\documentclass[12pt]{JHEP3}
\pdfoutput=1

\usepackage{picinpar}
\usepackage{epsfig}
\usepackage{amsmath}
\usepackage{amssymb}
\usepackage{cite}
\usepackage{graphicx}

\newcommand{\beqs}{\begin{equation*}}

\newcommand{\beq}{\begin{equation}}

\newcommand{\eeqs}{\end{equation*}}

\newcommand{\eeq}{\end{equation}}

\newcommand{\beqas}{\begin{eqnarray*}}

\newcommand{\beqa}{\begin{eqnarray}}

\newcommand{\eeqas}{\end{eqnarray*}}

\newcommand{\eeqa}{\end{eqnarray}}

\newcommand{\eq}[2]{\begin{equation} #1 \label{#2} \end{equation}}

\newcommand{\eps}{\varepsilon}

\newcommand{\al}{\alpha}

\newcommand{\ga}{\gamma}

\newcommand{\de}{\delta}

\newcommand{\om}{\omega}

\newcommand{\Ga}{\Gamma}

\newcommand{\De}{\Delta}

\newcommand{\Om}{\Omega}

\newcommand{\blist}{\begin{itemize}}

\newcommand{\elist}{\end{itemize}}

\providecommand{\href}[2]{#2}

\DeclareFontFamily{OT1}{rsfs}{}

\DeclareFontShape{OT1}{rsfs}{m}{n}{ <-7> rsfs5 <7-10> rsfs7 <10->rsfs10}{} 

\DeclareMathAlphabet{\mycal}{OT1}{rsfs}{m}{n}

\DeclareMathOperator{\extdm}{d}

\newcommand{\extd}{\extdm \!}

\newcommand{\tr}{{\rm tr}\;}

\newcommand{\vecX}{\boldsymbol{X}}
\newcommand{\vecJ}{\boldsymbol{J}}
\newcommand{\arctanh}{{\rm arctanh}}

\def\NO{\nonumber}

\def\bea{\begin{eqnarray}}
\def\eea{\end{eqnarray}}

\def\beqx{\begin{displaymath}}
\def\eeqx{\end{displaymath}}

\newcommand{\bmat}{\left(\begin{array}}
\newcommand{\emat}{\end{array}\right)}

\def\a{\alpha}
\def\b{\beta}
\def\d{\delta}
\def\e{\epsilon}
\def\f{\phi}

\def\h{\eta}

\def\l{\lambda}
\def\m{\mu}

    \def\om{\omega}
\def\p{\pi}

\def\r{\rho}
\def\s{\sigma}

\def\D{\Delta}

\def\G{\Gamma}

    \def\Om{\Omega}
\def\P{\Pi}

\def\ve{\varepsilon}

\def\vf{\varphi}

\def\cj{{\cal J}}

\def\co{{\cal O}}

\def\cq{{\cal Q}}

\def\cs{{\cal S}}

\def\cw{{\cal W}}

\def\bo{{\raise-.3ex\hbox{\large$\Box$}}}               
\def\pa{\partial}                                       
\def\face{{\raise.2ex\hbox{$\displaystyle \bigodot$}\mskip-2.2mu \llap {$\ddot
        \smile$}}}                                   
\def\>{\rangle}                                      
\def\<{\langle}                                      

\def\slash#1{\rlap{\hbox{$\mskip 1 mu /$}}#1}        
\def\leftrightarrowfill{$\mathsurround=0pt \mathord\leftarrow \mkern-6mu
        \cleaders\hbox{$\mkern-2mu \mathord- \mkern-2mu$}\hfill
        \mkern-6mu \mathord\rightarrow$}        
\def\dvec#1{\vbox{\ialign{##\crcr
        \leftrightarrowfill\crcr\noalign{\kern-1pt\nointerlineskip}
        $\hfil\displaystyle{#1}\hfil$\crcr}}}           
\def\tr{{\rm tr \,}}                                    

\def\-{\hphantom{-}}

\title{Bootstrapping gravity solutions}

\author{Jo\~{a}o Apar\'icio$^a$, Daniel Grumiller$^b$, Esperanza Lopez$^a$, Ioannis Papadimitriou$^a$ and Stefan Stricker$^b$\\
 $^a$~~Instituto de F\'{\i}sica Te\'orica UAM/CSIC, Universidad Aut\'onoma de Madrid,\\ 
 Calle Nicol\'as Cabrera 13--15, Madrid 28049, Spain\\
 $^b$~Institute for Theoretical Physics, 
          Vienna University of Technology,\\
          Wiedner Hauptstr. 8--10/136,
          A-1040 Vienna, Austria\\
Email: \email{jpmn.aparicio@gmail.com, grumil@hep.itp.tuwien.ac.at, esperanza.lopez@uam.es, Ioannis.Papadimitriou@csic.es, stricker@hep.itp.tuwien.ac.at}
}

\abstract{
We construct an algorithm to determine all stationary axi-symmetric solutions of 3-dimensional Einstein gravity with a minimally coupled self-interacting scalar field.
We holographically renormalize the theory and evaluate then the on-shell action as well as the stress tensor and scalar one-point functions.
We study thermodynamics, derive two universal formulas for the entropy and prove that global AdS provides a lower bound for the mass of certain solitons. 
Several examples are given in detail, including the first instance of locally asymptotically flat hairy black holes and novel asymptotically AdS solutions with non-Brown--Henneaux behavior.
}


\preprint{TUW--12--26\\ IFT UAM/CSIC--12--109}

\begin{document}

\section{Introduction}

Exact solutions have a long history in General Relativity, going back all the way to Schwarzschild in 1916, see \cite{Stephani:2003tm} and references therein.
Of particular interest are axi-symmetric stationary solutions: they have two commuting Killing vectors, both of which are phenomenologically viable symmetries in many physical situations, ranging from applications in astrophysics (black holes) to applications in the gauge/gravity correspondence (reasonable gravity duals for ground states or meta-stable states of the dual field theory).

Given some gravitational theory with matter --- for the sake of concreteness let us restrict to Einstein gravity --- it is notoriously difficult to derive all solutions to the equations of motion (EOM), which are non-linear coupled partial differential equations (PDEs). In 3+1 dimensions the imposition of axi-symmetry and stationarity still leads to quite complicated equations of that type. In 2+1 dimensions the PDEs are converted into ordinary ones, which is an essential simplification; however, even this simplification, which we implement in the present work, is in general not sufficient to permit an efficient algorithm to find all solutions.

It is much simpler, if not trivial, to find solutions in an inverse way: take some metric, declare it to be a solution of the Einstein equations with matter and deduce what kind of matter is needed for an exact solution. However, this procedure typically leads to solutions devoid of physical interest.

In this paper we propose a way to find all stationary axi-symmetric solutions to 2+1 dimensional Einstein gravity with a scalar field that simultaneously addresses the issue of physical relevance, similar in spirit to ``designer gravity'' \cite{Hertog:2004ns}. 

Our main ingredient is to avoid fixing the scalar self-interaction potential, but to extract it as an output. The crucial input, instead, is a certain function that determines essential properties of the geometry: its asymptotic behavior, the number and type of Killing horizons, the singularities etc. Thus, one can design the desired kind of geometry, while simultaneously guaranteeing the presence of reasonable matter --- a minimally coupled self-interacting scalar field that is regular, at least outside a possible black hole horizon. We call this procedure ``bootstrapping gravity solutions''.
As we shall demonstrate, our procedure is particularly useful in the context of the Anti-deSitter/Conformal Field Theory (AdS/CFT) correspondence \cite{Maldacena:1997re}. 
Let us state now some of our main results.
\begin{itemize}
 \item One key result is our solution generating algorithm presented in section \ref{se:2.1}, which requires one input function $f_1$ subject to an inequality, and several integration constants. Appropriate choices for the input function and integration constants are severely constrained by physics requirements (regularity, smoothness, suitable asymptotic behavior, absence of closed time-like curves, ...) discussed in section \ref{se:2.2}. 
 \item We find specific 2-parameter families of stationary black hole solutions in section \ref{se:2.3} and show that each such family has exactly one associated regular static solitonic solution. Moreover, we identify a constant of motion that generically leads to singular solutions when it is varied.
 \item In section \ref{se:3} we construct explicitly a family of locally asymptotically flat black hole solutions with scalar hair, with a symmetry-breaking scalar potential depicted in figure \ref{fig:1}. We also prove that there are no globally asymptotically flat black hole solutions with scalar hair.
 \item For asymptotically AdS boundary conditions (relaxing Brown--Henneaux) we holographically renormalize the theory in section \ref{se:BF}. We calculate the on-shell action and 1-point functions for Dirichlet, Neumann and Mixed boundary conditions for the scalar field. We focus particularly on marginal multitrace deformations, since in this case the 2-parameter black hole families that we construct can be interpreted as states of the same boundary theory.
 \item We generalize the results of section \ref{se:BF} and study the most general asymptotic AdS conditions compatible with locality of the holographically dual CFT in section \ref{se:4}. We specifically address subtleties that arise for resonant scalars and Breitenlohner--Freedman (BF) saturation. 
 \item We analyze the thermodynamics of black hole solutions in section \ref{se:TD} and prove two simple formulas for the entropy, one of which has a natural interpretation in terms of the Cardy formula, while the other one proves a 3-dimensional version of Penrose's Weyl curvature conjecture. Moreover, we discuss the thermodynamic (in)stability of black hole solutions and their associated solitons depending on the boundary conditions.
 \item In section \ref{se:5} we show how to recover known solutions within our algorithm and provide several new ones. We present in closed form two families of Brown--Henneaux solitons, with a potential displayed in figure \ref{fig:2}, as well as non-Brown--Henneaux solutions with oscillating potentials exhibited in figure \ref{Besselpot}. Since the most difficult step in our algorithm is the integration of a linear first order ordinary differential equation actually arbitrary solutions can be constructed, albeit in many cases one has to resort to (simple) numerics. In our set of examples we have cherry-picked those that can be presented in terms of fairly elementary functions.
\end{itemize}
In section \ref{se:6} we conclude with possible generalizations of our algorithm.

Before starting we mention some of our conventions. We use signature $(-,\,+,\,+)$ and fix Newton's constant by $16\pi G_N=1$.

\section{Einstein gravity with scalar field}\label{se:2}

Consider Einstein gravity with a minimally coupled self-interacting scalar field, to which we shall refer as Einstein-dilaton gravity (EDG). The EDG bulk action is
\eq{
S=\int\extd^3x\sqrt{-g}\,\Big[R+2-\frac12\,(\partial\phi)^2-V(\phi)\Big] 
}{eq:quench1}
where $V(\phi)$ is some (unspecified) self-interaction potential for the scalar field $\phi$. 
The potential is defined in such a way that $V(\phi)=0$ leads to a negative cosmological constant with unit AdS radius.

We are interested in stationary axi-symmetric solutions of the field equations descending from the action \eqref{eq:quench1}.
To this end we employ a method devised by Cl{\'e}ment \cite{Clement:1994sb}.
Namely, the line-element is parametrized as
\eq{
\extd s^2 = \lambda_{\mu\nu}\extd x^\mu \extd x^\nu + \frac{e^2\,\extd\rho^2}{-\det\lambda}
}{eq:quench3}
with the matrix $\lambda$ 
\eq{
\lambda_{\mu \nu} = \begin{pmatrix}
                     T+X & Y \\
                     Y   & T-X
                    \end{pmatrix}_{\mu\nu}
}{eq:quench4}
and all coefficient functions $T,X,Y$ as well as the Einbein $e$ only depend on the coordinate $\rho$.
It is convenient to introduce the Minkowskian target space vector $\vecX = (T,X,Y)$, whose norm is determined as $\vecX^2=-\det\lambda$.
The parametrization \eqref{eq:quench3} is accessible for any stationary axi-symmetric solution.
Inserting it into the action \eqref{eq:quench1} and dropping a boundary term yields the reduced action 
\eq{
S=\textrm{Vol}\,\int_{\rho_0}^\infty\limits\extd\rho\, e\,\Big[\frac{e^{-2}}{2}\,\dot\vecX^2 + 2 - \frac{e^{-2}}{2}\,\vecX^2\dot\phi^2 - V(\phi)\Big] 
}{eq:quench5}
where `$\textrm{Vol}$' refers to the volume of the 2-dimensional spacetime that was integrated out, dots denote $\extd/\!\extd\rho$ derivatives, and the integration domain is bounded by some $\rho=\rho_0$ and $\rho=\infty$. 
 
Solutions of the reduced action \eqref{eq:quench5} are in one-to-one correspondence to stationary axi-symmetric solutions of the original action \eqref{eq:quench1}.
After varying the reduced action with respect to the fields $e, \vecX, \phi$ we gauge-fix the Einbein to unity, $e=1$.\footnote{%
The Gaussian normal coordinate gauge employed in \cite{Deser:1986xf} is obtained for $e^2=-\det\lambda$.}
In this gauge the field equations read
\begin{subequations}
\label{eq:quench6} 
\begin{align}
& \textrm{Einstein:} && \ddot\vecX = -\dot\phi^2\vecX \label{eq:einstein} \\
& \textrm{Klein--Gordon:} && \vecX^2\ddot\phi + 2\vecX\dot\vecX \dot\phi = V'(\phi) \label{eq:kg} \\
& \textrm{Hamilton:} && -\frac12\,\dot\vecX^2+2+\frac12\,\vecX^2\dot\phi^2 - V(\phi) = 0 \label{eq:hamilton}
\end{align}
\end{subequations}
A first integral of the EOM is provided by the conserved particle angular-momentum (not to be confused with the black hole angular momentum of specific solutions),
\eq{
\eta^{nk}\,\epsilon_{ijk}\vecX^i\dot\vecX^j = \vecJ^n = \textrm{const.} 
}{eq:quench7}
The Klein--Gordon equation \eqref{eq:kg} can be reformulated as an equation for $\dot V$, which can then be integrated.
The integration constant is fixed by the Hamilton constraint \eqref{eq:hamilton}, which then leads to the following representation of the potential as a function of $\rho$.
\eq{
V(\rho) = 2 -\frac14\,\frac{\extd^2}{\extd\rho^2}\,\vecX^2
}{eq:quench8}

\subsection{Solution generating algorithm}\label{se:2.1}

The Einstein equation \eqref{eq:einstein} implies planarity of the vector $\vecX$.
Thus, with no loss of generality we parametrize generic solutions to the field equations \eqref{eq:quench6} as
\eq{
\vecX(\rho) = \vecX_1\,f_1(\rho) + \vecX_2\,f_2(\rho)
}{eq:quench9}
with some constant vectors $\vecX_{1,2}$.

Our bootstrapping algorithm to construct all solutions works as follows.
Pick some arbitrary function $f_1(\rho)$.
Then the first integral \eqref{eq:quench7} dictates that $f_2$ be a solution of 
\eq{
\dot f_2 f_1 - \dot f_1 f_2 = 2j = \rm const.
}{eq:quench12}
The constant $j$ enters the angular momentum in \eqref{eq:quench7}.
The first order ODE \eqref{eq:quench12} can be interpreted as Wronskian. 
Consequently, two linearly dependent functions $f_1\propto f_2$ always require vanishing $j$ (and thus vanishing $\vecJ$), while non-vanishing $j$ implies their linear independence.

The action \eqref{eq:quench5} is Lorentz invariant, a property inherited from the $SL(2,\mathbb{R})$ invariance of the parameterization \eqref{eq:quench3}, which we can use to bring the constant vector $\vecJ$ into some convenient form. For non-vanishing $j$ there are three distinct cases: time-like, light-like or space-like $\vecJ$. For reasons that will become clear later we focus almost exclusively on the case when $\vecJ$ is space-like in our work.
Then it is always possible to find Lorentz transformations that simultaneously set to zero the $Y$-component in both vectors $\vecX_{1,2}$.
Thus, for space-like $\vecJ=(0,\,0,\,-j)$ with no loss of generality we rewrite the parametrization \eqref{eq:quench9} as
\eq{
\vecX(\rho) = \begin{pmatrix}
               T\\
               X\\
               Y
              \end{pmatrix} = \frac12\,\begin{pmatrix}
               f_1 - f_2\\
               -f_1 - f_2\\
               0
              \end{pmatrix}
}{eq:quench10}
which implies that the 2-dimensional metric $\lambda_{\mu\nu}$ in \eqref{eq:quench4} simplifies to
\eq{
\lambda_{\mu\nu} = \begin{pmatrix}
               - f_2 & 0 \\
               0 & f_1 
              \end{pmatrix}_{\mu\nu}\,.
}{eq:quench11}
Another consequence of the Einstein equations is that the functions $f_{1,2}(\rho)$ are not independent from each other:
Both of them are solutions of the second order ODE $\ddot f_{1,2} + \dot\phi^2f_{1,2}=0$.

The next step determines the scalar field $\phi$ as a function of $\rho$ by virtue of the Einstein equation \eqref{eq:einstein}.
(For later convenience we fix a sign ambiguity by assuming that $\dot\phi$ is negative.)
\eq{
\dot\phi = -\sqrt{-\frac{\ddot f_1}{f_1}} = -\sqrt{-\frac{\ddot f_2}{f_2}}
}{eq:quench13}
Note that a reality condition on the scalar field $\phi$ implies a convexity/concavity property for the function $f_1$.
To be precise, $f_1$ has to be concave for positive values of $f_1$ and convex for negative values of $f_1$.
Thus, not any choice of the function $f_1$ leads to a suitable solution.
Rather, we have to demand that in the range of definition of $\rho$ the following inequality is valid.
\eq{
\frac{\ddot f_1}{f_1} \leq 0
}{eq:quench14}
This implies that $f_1$ cannot have any divergence at finite $\rho$. The same is true for $f_2$.

The next-to-last step is purely algebraic: 
The field equation \eqref{eq:quench8} determines the potential $V$ as a function of $\rho$.
\eq{
V = 2 - \frac12\, \frac{\extd}{\extd\rho} \big(f_1\dot f_2\big)
}{eq:quench15}
At this stage we know $\phi$ and $V$ as functions of $\rho$.
Since $\phi(\rho)$, when it is non-constant, is strictly monotonous within the range of definition we can, at least in principle, uniquely determine $\rho(\phi)$.
Thus, implicitly we know also the potential $V(\phi)$.

With the algorithm so far we can obtain all local solutions by scanning through all input functions $f_1$ compatible with the inequality \eqref{eq:quench14}. The last step is to relate the coordinates $x^0$ and $x^1$ used in the local construction above to time $t$ and angular coordinate $\varphi$ defined by their periodicity properties $(t,\,\varphi)\sim(t,\,\varphi+2\pi)$. They must be linear combinations of the coordinates $(x^0,\,x^1)$, so that 
\eq{
t = a x^0 + b x^1 \qquad \varphi = c x^0 +d x^1
}{eq:quench137}
Clearly, the above transformations generate solutions of the EOM with the same scalar potential.
General solutions will then have the form
\eq{
\extd s^2 = -f_2(\rho) (\extd x^0)^2 + f_1(\rho) (\extd x^1)^2 + \frac{\extd\rho^2}{f_1(\rho)f_2(\rho)}
}{eq:quench18}
identifying the angular and time coordinates according to \eqref{eq:quench137}.
For static spacetimes we fix $a=d=1$ and $b=c=0$ so that $\varphi = x^1$ and $t = x^0$
\eq{
\extd s^2 = -f_2(\rho) \extd t^2 + f_1(\rho) \extd\varphi^2 + \frac{\extd\rho^2}{f_1(\rho)f_2(\rho)}
}{eq:quench19}

Some useful geometric quantities are the Ricci scalar
\eq{
R = - f_1\ddot f_2 - f_2 \ddot f_1 - \frac32\,\dot f_1 \dot f_2 = 2 f_1 f_2\,\dot\phi^2- \frac32\,\dot f_1 \dot f_2
}{eq:quench25}
the square of the tracefree Ricci tensor $\slash{R}_{\mu\nu}=R_{\mu\nu}-\tfrac13\,g_{\mu\nu}R$
\eq{
\slash{R}_{\mu\nu} \slash{R}^{\mu\nu} = \frac16\,f_1^2f_2^2\,\dot\phi^4
}{eq:quench21}
and the Cotton tensor (see e.g.~\cite{Garcia:2003bw} for a definition)
\eq{
C_{\mu\nu} = \frac j4 \, f_1 f_2\,\dot\phi^2 \, \big( \de_\mu^0\de_\nu^1 + \de_\mu^1 \de_\nu^0\big)\,.
}{eq:quench26}
All solutions with $j=0$ or constant scalar field, $\dot\phi=0$, are conformally flat.

We finally note that a given solution determined by the functions $f_{1,2}$ permits to construct new solutions by taking linear combinations
\eq{
\tilde f_1 = A f_1 + B f_2 \qquad 
\tilde f_2 = C f_1 + D f_2
}{eq:quench54}
where $A, B, C, D$ are some real numbers.
Feeding the functions $\tilde f_{1,2}$ into our algorithm then leads to new solutions analog to \eqref{eq:quench18}, since \eqref{eq:quench12} is invariant under \eqref{eq:quench54}, with $\tilde\jmath=j\,(AD-BC)$.
All these solutions have the same scalar field \eqref{eq:quench13}, but in general not the same potential \eqref{eq:quench15}. Thus, these solutions are in general not solutions of the same theory. A notable exception is $A=D=0$, $B=C=1$, which we shall use in section \ref{se:2.3} to construct solitons from black hole solutions.


From the solutions \eqref{eq:quench18}, \eqref{eq:quench19} it is clear that the function $f_{1,2}$ should be chosen conveniently, depending on the application. ``Conveniently chosen'' refers to the sub-sector of solutions we are interested in --- for instance, we may wish to restrict to asymptotic AdS solutions or to asymptotically flat solutions, to solutions with or without horizons, etc. Moreover, certain conditions have to be imposed on the functions $f_{1,2}$ to guarantee the absence of naked singularities. Imposing all these conditions from the very beginning of the algorithm is an important part of it, as this restricts the possible choices for the function $f_1$ and the integration constants appearing in the function $f_2$.

\subsection{General features --- asymptotics, singularities, horizons, and centers}\label{se:2.2}

In this section we analyze how to engineer solutions of EDG with the desired properties. 
We assume throughout in this paper that spacetime is non-compact and that the asymptotic region is reached in the limit $\rho\to\infty$.
This means that we pre-select in all our constructions the functions $f_{1,2}$ such that this property holds.
Moreover, we assume in most of our discussion that $\vecJ$ is space-like. The case of time-like and light-like $\vecJ$ is addressed at the end of this subsection.

If we move from the asymptotic region towards the interior by considering smaller values of the ``radial'' coordinate $\rho$ then eventually we may hit a singularity (curvature singularity, conical singularity, region with closed time-like curves or more exotic singularities);
alternatively, we might not encounter any singularity but continue until we reach another asymptotic region in the limit $\rho\to-\infty$.
For most applications we are not interested in either of these possibilities.
Instead, we would like spacetime to either contain a Killing horizon, shielding possible singularities (or superfluous asymptotic regions), or to contain a center, by which we mean a region around which spacetime looks locally like flat spacetime in polar coordinates close to the origin.
In the former case an outside observer has access to the exterior region outside the Killing horizon, $\rho\geq \rho_h$, while in the latter case the whole spacetime is restricted to the semi-infinite interval $\rho\in[\rho_0,\infty)$, where $\rho=\rho_0$ is the locus of the center.
It is therefore of interest to establish a necessary condition on the functions $f_{1,2}$ for the emergence of a Killing horizon or a center.

\paragraph{Restrictions from regularity}
We consider the restrictions that smoothness around a chosen value of the radial coordinate imposes on solutions of the EOM.
Assuming that the leading behavior at $\rho=\bar\rho$ is power law, two linearly independent functions satisfying  \eqref{eq:quench12} are
\eq{
f \propto (\rho -{\bar \rho})^\alpha  \qquad {\tilde f} \propto (\rho -{\bar \rho})^{1-\alpha}   
}{eq:add1}
where $\alpha\!>\!1/2$  and we have ignored subleading terms at $\rho\!\sim\!{\bar \rho}$. [For $\alpha\!=\!1/2$, ${\tilde f}\!\propto\!f \log(\rho-{\bar \rho})$.]
From \eqref{eq:quench13},  a pair $f_1$, $f_2$ constructed from linear combinations of the above two functions gives rise to the following behavior for the scalar field 
\eq{
\dot\phi = -{ \sqrt{\alpha(1-\alpha)}\over | \rho-{\bar \rho}|} 
}{eq:add2}
Reality of the scalar around ${\bar \rho}$ requires $\alpha\!\leq\!1$. Besides, unless $\alpha\!=\!1$ 
the scalar field develops a logarithmic singularity and, as can be seen from \eqref{eq:quench25}-\eqref{eq:quench26}, there is a curvature singularity at the same locus.
Having $\alpha$ equal unity is however not enough for smoothness at $\bar \rho$. According to \eqref{eq:quench13} we also need to demand
\eq{
\ddot f({\bar \rho}) = 0\,.
}{eq:add22}
If this condition is not satisfied, including the first subleading term at ${\bar \rho}$ we have
\eq{
f \propto (\rho-{\bar \rho})+{1 \over 2} \ddot f({\bar \rho}) ( \rho-{\bar \rho})^2 \qquad {\tilde f} \propto 1+\ddot f({\bar \rho})( \rho-{\bar \rho}) \log( \rho-{\bar \rho})
}{eq:add222}
which induces a curvature singularity (unless $f_1\propto f_2 \propto f$).

A similar analysis can be done in the asymptotic region. Regularity of the scalar field at infinity implies that $f_1$, $f_2$ are linear combinations of functions whose leading behavior at large values of the radial coordinate is $f \!\propto \!\rho$ and ${\tilde f}\! \propto \!1$. In the following we shall restrict to this case. 
It leads to a constant value of the Ricci scalar at infinity
\eq{
R \rightarrow -{3 \over 2} {\dot f}_1{\dot f}_2
}{eq:add23c}
and a vanishing Cotton tensor. Moreover, the metric asymptotes locally to (A)dS when $f_1$ and $f_2$ grow linearly and the relative sign of both functions is (equal) opposite. 
If either $f_1$ or $f_2$ tends to a constant, the curvature vanishes at large $\rho$ and the spacetime is locally asymptotically flat. 

\paragraph{Zeroes of the functions $f_{1,2}$}

Without loss of generality we choose $f_1$ to be asymptotically positive.
Taking $f_1$ as input in our algorithm, we obtain $f_2$ integrating \eqref{eq:quench12}
\eq{
f_2= \chi f_1 - 2j f_1 \int\limits_\rho {\extd \rho' \over f_1^2} 
}{eq:add3}
where $\chi$ and $j$ are integration constants of the EOM, $j$ having the interpretation of particle angular momentum in the reduced model \eqref{eq:quench5}. The second term on the right hand side of \eqref{eq:add3} provides a second solution, linearly independent from $f_1$, to \eqref{eq:quench12}. If $f_1$ grows linearly at large values of the radial coordinate, this second term tends to a constant and vice versa. 

The inequality \eqref{eq:quench14} implies that $f_1$ has a zero at a finite value of the radial coordinate, $\rho_1$, if we assume regularity of $f_1$ for $\rho_1\!\leq\!\rho\!<\!\infty$ at least up to its second derivative. We want to determine under which conditions $f_2$ can have its first zero at $\rho_2\!>\!\rho_1$. This is equivalent to search for the vanishing of
\eq{
\chi - 2j  \int\limits^R_\rho {\extd \rho' \over f_1^2}
}{eq:add4}
where we have introduced the upper integration limit $R$ for concreteness; its only effect is to shift the value of the constant $\chi$. 
The behavior of the integral at $\rho_1$ depends on how fast $f_1$ approaches zero there. Let us consider first that it vanishes sufficiently fast to induce a divergence. 
The integral is a monotonously decreasing function. When $f_1\!\propto\!\rho$ at large values of the radial coordinate, it is convenient to set $R\!=\!\infty$ such that the integral vanishes asymptotically. Hence \eqref{eq:add4} will have one and only one zero for any $j/\chi>\!0$. By construction $\rho_2$ is a single zero of $f_2$ with vanishing second derivative. 

The sign of $\chi$ determines whether the spacetime approaches AdS or dS asymptotically. 
When $\chi\!=\!0$ the spacetime is asymptotically flat, but there is no value of the radial coordinate where \eqref{eq:add4} becomes zero.
An asymptotically flat spacetime is also obtained for $f_1$ approaching a constant at infinity. In this case the integral in \eqref{eq:add4} diverges at large $\rho$ to minus infinity,
and that expression has one and only one zero for any $j$ and $\chi$. 

When instead the function $f_1$ does not vanish fast enough at $\rho_1$ to generate a divergent contribution to the integral, there will be a maximum value of $j/\chi$ for which \eqref{eq:add4} can have a zero at $\rho_2\!\geq\!\rho_1$. At the limiting value the zero occurs precisely at $\rho_1$. In the previous case, as $j/\chi\!\rightarrow\!0^+$ we have $\rho_2\!\rightarrow\! \rho_1$.  

Let us summarize our results. Assuming the regularity of $f_1$ up to its second derivative and $\rho_1\!<\!\rho_2$, the function $f_2$ has a regular zero at the point $\rho_2$: a simple zero with vanishing second derivative. If $\rho_1$ is also a regular zero of $f_1$ the solution can be prolonged beyond this point, otherwise the  lowest value of the radial coordinate where the solution is defined is $\rho_0\!=\!\rho_1$.
Let us stress the importance of the regularity condition on $f_1$. If $f_2$ has a simple zero with non vanishing second derivative, the function $f_1$ is finite but not differentiable at that point, see \eqref{eq:add222}. As we have discussed this induces a curvature singularity, implying $\rho_0\!=\!\rho_2$. Similar conclusions hold when $\rho_2\!<\!\rho_1$. 
Regular zeroes of the functions $f_1$ and $f_2$ must alternate. [This as can be seen in the very simple example $f_1\!=\!\cos \rho$, $f_2\!=\!\sin \rho$.] 
When $\rho_1\!=\!\rho_2$ and $j\!\neq\!0$, necessarily there is a curvature singularity at that locus and then $\rho_0=\rho_{1,2}$.

\paragraph{ADM gauge fixing}
Let us write the line-element \eqref{eq:quench18} explicitly in terms of the time and angular coordinates
\eq{
\extd s^2 = -N^2 \extd t^2 + \frac{\extd \rho^2}{N^2 R^2} + R^2 \big(\extd \varphi + N_\varphi \extd t\big)^2
}{eq:ADM}
where
\eq{
R^2 = a^2 f_1 - b^2 f_2 \qquad N^2 = {f_1f_2 \over R^2} \qquad N_\varphi  = \frac{ bd  f_2-ac f_1}{R^2}
}{eq:add00}
Regions with $R^2$ negative contain closed time-like curves. Hence we shall always assume $R^2$ to be positive in the asymptotic region. Defining a new radial variable as $\extd r^2 = \extd\rho^2/R^2$, the usual ADM form of the metric is recovered, with $N$ being the lapse function and $N_\varphi$ contributing to the shift vector.

The constants $a,b,c,d$ were introduced in \eqref{eq:quench137}. The coordinate change \eqref{eq:quench137} with the restriction $ad\!-\!bc\!=\!1$ provides a realization of the $SL(2,\mathbb{R})$ group of transformation of the matrix $\lambda_{\mu \nu}$, which locally preserves the line-element \eqref{eq:quench3}. A global rescaling of $a,b,c,d$ implements the multiplication of $\lambda_{\mu \nu}$ by a constant that, together with a compensating rescaling of the radial coordinate, also preserves Cl\'ement's parameterization of the line-element. It will be convenient however to implement global rescaling directly on $\vecX$, while the $SL(2,\mathbb{R})$ group is realized in terms of the map between local and physical variables \eqref{eq:quench137}. This choice has been already implemented in \eqref{eq:ADM}, where $a,b,c,d$ are taken to satisfy the $SL(2)$ condition. In order to avoid redundancies in the local description, we need then to select a representative for each $SL(2,\mathbb{R})$ orbit of the vector $\vecX$. This is partly achieved with the gauge choice $Y\!=\!0$, but it still allows for the family of solutions $(\ga f_1,f_2/\ga)$ with $\ga \!\in \!\mathbb{R}$. We will fix this remaining $SL(2,\mathbb{R})$ freedom as follows. When both $f_{1,2}$ grow linearly at infinity, we require that $f_1\!\rightarrow\!\pm f_2$ asymptotically. When $f_1$ or $f_2$ tend asymptotically to a constant, we will set this constant to $\pm 1$.

\paragraph{Horizons}
A sufficient condition for the existence of a Killing horizon at a point $\rho_h$ of finite curvature is the vanishing of the lapse function $N$ at that point. 
From \eqref{eq:add00} and the discussion above, this implies that $f_1$, $f_2$ or both must have a regular zero at $\rho_h$. 

When $j\!\neq\!0$, the functions $f_{1,2}$ are linearly independent and hence only one of them will vanish at $\rho_h$. Let us assume that $R^2>0$ at $\rho_h$, so that we are not in a region of closed time-like curves.
Reparametrizing the radial direction as $y = \sqrt{2(\rho-\rho_h)/|j|}$ close to the zero, the line-element simplifies to
\eq{
\extd s^2 = -\epsilon\, {j^2 \over R^2_h} \, y^2 \extd t^2 + \epsilon  \extd y^2 + R^2_h \, ( \,\extd\varphi + N_{\varphi \, h} \extd t \,)^2  + \dots
}{eq:quench61}
where the subindex `$h$' indicates that the corresponding function is evaluated at $\rho_h$. We have defined $\epsilon\!=\!(-1)^{i} \, {\rm sign}(j)$ with $i\!=\!1$($2$) for vanishing $f_1$($f_2$).  When $\epsilon\!=\!1$ the first two terms in \eqref{eq:quench61} yield 2-dimensional Rindler spacetime, which is known to be the correct near horizon approximation of any non-extremal black hole Killing horizon (see e.g.~\cite{waldgeneral}). When $\epsilon\!=\!-1$ we obtain instead an inner black hole horizon or cosmological horizon. A consequence of \eqref{eq:quench26} is that all Killing horizons have locally vanishing Cotton tensor.

When $j\!=\!0$, condition \eqref{eq:quench12} implies the proportionality of $f_{1,2}$. Therefore in this case both $f_{1,2}$ vanish at a Killing horizon. Using then the coordinate transformation $y\!=\!\sqrt{{\dot f} (\rho-\rho_h)}$ we obtain
\eq{
\extd s^2 = - y^2 \extd t^2 + y^2 \,\big(\extd\varphi + N_\varphi\extd t\big)^2 + {4 \over \dot f^2}\, {\extd y^2 \over y^2} + \dots
}{eq:quench63}
where according to our previous gauge choice $f_{1,2}\!=\! f$.
The quantity $N_\varphi$ is constant for vanishing $j$.
For $N_\varphi=0$ the metric \eqref{eq:quench63} describes the Poincar\'e patch horizon of AdS.

Extremal black hole horizons are somewhat delicate with our gauge choices. A straightforward way to obtain them is from a 2-horizon configuration, performing suitable $SL(2)$ transformations, and taking the limit where both horizons coincide in the end. In the near horizon approximation we have $f_1=2(\rho-\rho_+)+2\eps\rho_+$, $f_2=2(\rho-\rho_+)$,  where $\eps>0$ is a small parameter that we send to zero eventually. We assumed here that at $\rho=\rho_+>0$ we have a black hole horizon [$f_2(\rho_+)=0$] and at some slightly smaller value an inner horizon [$f_1(\rho_+-\eps\rho_+)=0$]. Our first $SL(2)$ transformation rescales the functions by (one over) $\eps$, $f_1=2(\rho-\rho_+)/\eps+2\rho_+$, $f_2=2(\rho-\rho_+)\eps$. Our second $SL(2)$ transformation is generated by \eqref{eq:quench137} with $a=c=1$, $b=1/\eps-1/2$, $d=1/\eps+1/2$. Note that this transformation is singular at $\eps=0$, which is why so far we have to work at finite $\eps$. However, after these transformations we can smoothly take the limit $\eps\to 0$ and obtain the near-horizon line-element
\eq{
\extd s^2 = -\frac{\extd t^2 2(\rho-\rho_+)^2}{\rho} + \frac{\extd\rho^2}{4(\rho-\rho_+)^2} + 2\rho\,\big(\extd\varphi-\frac{\rho_+}{\rho}\,\extd t\big)^2\,.
}{eq:extremalBH}
Introducing $r=\sqrt{2\rho}$ leads to the extremal BTZ black hole in one of the standard coordinate systems, which is the near horizon limit of any extremal black hole in EDG. Note that close to $\rho=\rho_+$ the first two terms in the line-element \eqref{eq:extremalBH} form the expected AdS$_2$ factor, see for instance \cite{Strominger:1998yg}. 

\paragraph{Global restrictions}
We consider now to what extent the existence of a Killing horizon is compatible with different asymptotic behaviors. As before we assume $f_1$ to be asymptotically positive, so the different cases depend only on the asymptotic sign of the function $f_2$, $\sigma=\lim_{\rho\to\infty}\textrm{sign}\,(f_2)$. As discussed above \eqref{eq:add23c}, $\sigma$ cannot be zero for asymptotically regular solutions, so we need to discuss only two cases.
\begin{itemize}
\item[$\s=+$] This covers asymptotically locally AdS and flat spacetimes. In order to have $R^2$ positive at the outermost horizon, the function $f_2$ must vanish there, implying $\epsilon\!=\!1$. Behind the horizon we may encounter: {\it i)} a curvature singularity where $f_{1,2}$ vanish; {\it ii)} a ring of curvature singularities of the type \eqref{eq:add222}, where only $f_1$ vanishes while $R^2$ is positive; {\it iii)} a regular zero of $f_1$ with $\epsilon\!=\!-1$ producing an inner Killing horizon. 
Notice that an inner horizon is only present in the non-generic situation {\it iii)}, otherwise there is a curvature singularity. This agrees with the results of \cite{Banados:1992gq}, obtained studying matter fluctuations around a BTZ black hole. For static spacetimes, the ring of singularities in case {\it ii)} collapses to a point, while the regular zero of $f_1$ in case {\it iii)} gives rise to a Milne-type singularity (a Minkowski version of a conical singularity).
\item[$\s=-$] This covers asymptotically locally dS and asymptotically locally flat cosmological solutions. The outermost Killing horizon necessarily satisfies $\epsilon\!=\!-1$, and consistently has the interpretation of a cosmological horizon. Inside the cosmological horizon it turns out not to be possible to find a second Killing horizon without reaching before a region with $R^2$ negative and thus closed time-like curves. Hence this rules out the existence of regular black hole geometries in asymptotically dS spacetimes in the context of EDG.  
\end{itemize}
We will not have more to say in this paper about asymptotically dS or flat cosmological solutions (see \cite{Bagchi:2012yk}). 
Hence from now on we focus on asymptotically positive functions $f_{1,2}$ as input for our algorithm.

\paragraph{Centers} 
A necessary condition for the existence of a center at $\rho_0$ is the vanishing of the radial function $R$ while the lapse function remains finite.
This can only be achieved  if either $a$ or $b$ (but not both) vanish. Requiring $f_{1,2}$ and $R^2$ to be asymptotically positive necessarily selects $b\!=\!0$.
Hence centers are associated with regular zeroes of $f_1$. 
To reduce clutter we focus on static spacetimes. (Stationary spacetimes can be treated analogously.)
Changing again the radial coordinate to $y = \sqrt{2 (\rho-\rho_0)/|j|}$, the line-element close to such a zero simplifies to
\eq{
\extd s^2 = -f_{2\,c} \extd t^2 + {j^2 \over f_{2 \,c}} \,y^2\extd\varphi^2 + \extd y^2 + \dots
}{eq:quench62}
with $f_{2\,c}\!=\!f_2(\rho_0)$. Smoothness requires furthermore the condition 
\eq{
f_{2\,c}=j^2\,.
}{eq:add0}
If this condition is violated there is a conical defect at $\rho=\rho_0$.

Notice that the relation \eqref{eq:add0} together with positivity of $\dot f_{1\,c}$ and the condition \eqref{eq:quench12} imply that $j$ must be negative. 
By contrast, static solutions with a horizon and a regular zero of $f_1$ discussed in the previous paragraph have positive $j$. 

\paragraph{Light-like or time-like $\vecJ$} If the angular momentum vector $\vecJ$ in  \eqref{eq:quench7} is not space-like (and not zero) then the following inequality holds.
\eq{
\det\lambda \leq 0
}{eq:referee1}
If $\vecJ$ is light-like, the inequality \eqref{eq:referee1} can be saturated, while for time-like $\vecJ$ the inequality is strict. This implies that there can be no region of spacetime with positive $\det\lambda$, so that the radial coordinate $\rho$ in \eqref{eq:quench3} cannot be time-like anywhere. Therefore, non-extremal Killing horizons do not exist in these cases. In fact, for time-like $\vecJ$ no Killing horizon at all can be present, so that we neither can have black holes nor cosmological horizons. This is the reason why we focussed our discussion on the more interesting case of space-like $\vecJ$, where all these possibilities can arise.

\newcommand{\matr}{M_{R^2}}
\newcommand{\matn}{M_{N_\varphi}}

In order to prove that zeros of $\det\lambda$ are necessary for the appearance of Killing horizons we generalize now the results for lapse, shift and radial function \eqref{eq:add00} for arbitrary matrices $\lambda$,
\eq{
R^2=\tr(\lambda \matr)\qquad N^2=-\frac{\det\lambda}{R^2}\qquad N_\varphi = \frac{\tr(\lambda \matn)}{R^2}
}{eq:referee2}
with the rank-1 matrices
\eq{
\matr = \begin{pmatrix}
     b^2 & -ab \\ -ab & a^2
    \end{pmatrix}
\qquad
\matn = \begin{pmatrix}
     -bd & bc \\ ad & -ac
    \end{pmatrix}\,.
}{eq:referee3}
Since $R^2$ must be finite on Killing horizons, the condition of vanishing lapse function, $N^2=0$, is equivalent to the condition of vanishing determinant, $\det\lambda=0$.

No static solutions exist for time-like or light-like $\vecJ$, since all allowed solutions of $N_\varphi=0$ require the functions $f_{1,2}$ in \eqref{eq:quench9} to be proportional to each other. But this means that $\vecX$ and $\dot\vecX$ depend linearly on each other and therefore the angular momentum vector \eqref{eq:quench7} vanishes. Similarly, no regular centers exist, since solutions of $R^2=0$ do not lead to finite lapse functions. Therefore, there can also be no regions of closed time-like curves, $R^2<0$, unless the whole spacetime exhibits them.

The only interesting global aspects for time-like angular momentum vectors $\vecJ$ are then the asymptotic structure and the possible presence of curvature singularities. Their discussion is analog to the space-like case. Note that curvature singularities are necessarily naked. 

Light-like angular momentum vector $\vecJ$ is more interesting than time-like one, since it can lead to extremal horizons. This is so, because the inequality \eqref{eq:referee1} can be saturated. Whenever this happens $\det\lambda$ has an even zero (e.g.~a double zero). Thus, if one wants to avoid the delicate scaling limit explained above to describe extremal horizons one can simply start with light-like $\vecJ$. For instance, starting with the matrix
\eq{
\lambda_{\mu\nu} = \begin{pmatrix}
           f_1 & f_2 \\ f_2 & 0
          \end{pmatrix}_{\mu\nu}
}{eq:referee4}
we obtain $\vecJ=-j\,(1,\,1,\,0)$. If $f_2$ has a single zero, $f_2\propto (\rho-\rho_+)+ \dots$, the near-horizon line-element is precisely of the form \eqref{eq:extremalBH}.

For the above reasons --- no static solutions, no solutions with center, no solutions with non-extremal Killing horizon in the case of time-like or light-like $\vecJ$ --- in the remainder of the paper we focus on space-like $\vecJ$.

\subsection{Constants of motion vs.~parameters of the model}\label{se:2.3}

The algorithm above involves a number of free constants and provides a potential $V$ as an output that depends in general on these constants.
However, once a potential $V$ is constructed it is also of interest to obtain further solutions to the EOM with the {\em same} potential. 
These new solutions differ from the previous one by the values of certain constants of motion.

\paragraph{Counting of constants of motion}
The Einstein equations \eqref{eq:einstein} and Klein--Gordon equations \eqref{eq:kg} lead to eight integration constants.
The Hamilton constraint \eqref{eq:hamilton} relates them, so that we have seven independent constants of motion. 
(Time derivatives of the Hamilton constraint do not lead to new relations between these constants of motion.)
Three of these constants are contained in the $SL(2,\mathbb{R})$ transformations of the matrix $\lambda_{\mu\nu}$ in \eqref{eq:quench4}.
Actually, two of the $SL(2,\mathbb{R})$ transformations have been used to eliminate the variable $Y$ in the matrix \eqref{eq:quench4}.
Since the action \eqref{eq:quench5} does not depend explicitly on the radial coordinate, shifts of $\rho$ generate new solutions. This freedom can be parameterized by the value of $\rho_0$. Another integration constant is the conserved particle angular momentum parameter $j$.
The remaining two constants of motion, denoted by $\phi_\pm$, are coefficients of the two scalar modes and appear implicitly in the specification of the function $f_1$, the starting point of our algorithm. We shall see explicitly the emergence of these constants when discussing asymptotically AdS solutions in sections \ref{se:BF} and \ref{se:4}.

\paragraph{Co-dimension 2 family of solutions}
Contrary to $SL(2,\mathbb{R})$ transformations and radial shifts, our algorithm does not instruct us how solutions must depend on $j$ and $\phi_\pm$ in order for the scalar potential to be independent of them.
Given $f_1$ we can construct $f_2$ for any value of $j$ from \eqref{eq:add3}. But these solutions will not relate to the same potential unless $f_1$ carries an a priori involved dependence on $j$, which we do not know how to implement systematically. There is, however, a simple way of constructing a one parameter family of solutions to the same $V(\phi)$ which moves in the interesting subspace of constants of motion $\{j,\phi_\pm\}$. It is generated by the rescaling of $f_{1,2}$ together with a compensating rescaling of the radial coordinate
\eq{
f_{1,2}(\rho) \to \Om f_{1,2} (\rho/\Om)\qquad \phi(\rho) \to\phi(\rho/\Om)
}{eq:quench147}
Here $f_1$ and $f_2$ are assumed to solve the EOM for some $j$ and $\phi_\pm$. Then
\eq{
j \to \Om j\qquad \phi_\pm \to \Om^{\De_\pm/2}\phi_\pm
}{eq:quench148} 
where $\De_\pm$ are the scaling dimensions of $\phi_\pm$. (Also $\rho_0 \to \Om \rho_0$.) As already mentioned, transformation \eqref{eq:quench147} is a symmetry of Cl\'ement's parametrization of the line-element, and hence of the particle model \eqref{eq:quench5}. This guarantees the scalar potential to be $\Om$-independent. Alternatively, the fact that both the scalar field and the potential as function of the radial coordinate \eqref{eq:quench15} depend on $\rho$ and $\Om$ through $\rho/\Om$, leads to the same conclusion.
Motivated by the convenience to use $\Om$ as one of the constants of motion, we change parametrization in the space of solutions from $\{j,\phi_\pm\}$ to $\{\Om,\xi,\lambda\}$, where 
\eq{
\xi=j/\Om \qquad  \lambda= \phi_+/ \phi_-^{\De_+/\De_-} 
}{eq:add5}

\paragraph{Analytic expansions}
As implied by \eqref{eq:quench13}, smooth solutions with a Killing horizon or a center require $f_{1,2}$ to be at least twice differentiable down to its locus. In order to carry on a general analysis, we further assume now all functions to be analytic for $\rho\!>\!\rho_0$. Around a generic radial point $\bar \rho$, we have
\eq{
f_1 =  \sum_{n=0}^\infty a_n {(\rho-{\bar \rho})^n\over \Om^{n-1}}\qquad f_2=\sum_{n=0}^\infty b_n {(\rho-{\bar \rho})^n\over \Om^{n-1}}\qquad \phi=\sum_{n=0}^\infty \phi_n {(\rho-{\bar \rho})^n\over \Om^n}
}{eq:quench149}
The conservation law \eqref{eq:quench12} implies
\begin{align}
 & a_0 b_1-a_1b_0=2\xi \label{eq:add6} \\
 & a_0 b_2-a_2b_0=0 \label{eq:add7} \\[-2mm]
 & \dots \nonumber
\end{align}
which fixes the coefficients $b_{n>0}$ in terms of $\xi$, $a_{m<n}$ and $b_0$. Similarly, the defining equation for the scalar field \eqref{eq:quench13} determines the coefficients $\phi_{n>0}$ in terms of $a_{m<n+2}$. The potential as a function of the radial coordinate is given by
\eq{
V = 2 -\frac14\,\sum_{n=2}^\infty n(n-1)\sum_{m=0}^n a_m b_{n-m} {(\rho-{\bar \rho})^{n-2}\over \Om^{n-2}}
}{eq:quench150}
On the other hand, assuming that the potential $V(\phi)$ is also an analytic function of $\phi$ around $\phi({\bar \rho})\!=\!c_0$ yields
\eq{
V = V(\phi_0) + (\phi-\phi_0) V^\prime(\phi_0) + \frac{(\phi-\phi_0)^2}{2}\,V''(\phi_0) + \dots
}{eq:quench151}
Equating the two expansions \eqref{eq:quench150} and \eqref{eq:quench151} we obtain to leading order
\eq{
V(\phi_0) = 2 - \frac12\big(a_0 b_2 + a_1b_1+a_2 b_0\big)
}{eq:quench152} 
This relation determines the coefficient $a_2$ in terms $a_0,b_0,a_1,\phi_0$ and $\xi$.
Comparison to higher orders fixes $a_{n>2}$ as functions of the same parameters. 
The parameters $a_0$, $b_0$ and $\Omega$ are not independent from each other; instead, one of them can be fixed in terms of the others.
Consistently, we recover that the space of solutions of EDG in the gauge $Y\!=\!0$ is five-dimensional.

\paragraph{Static black holes and their associated solitons}
We focus now on the interesting case of having a black hole horizon at $\bar \rho$. As discussed in the previous section, this requires to impose $b_0\!=\!0$, while the regularity condition \eqref{eq:add22}, viz.~$\!b_2\!=\!0$, is automatically satisfied by \eqref{eq:add7}. Using the freedom to redefine $\Om$ we can set the product $a_0b_1$ to any desired value. For convenience we choose 
\eq{
a_0 b_1=8\,.
}{eq:add200}
Relation \eqref{eq:add6} fixes then $\xi\!=\!4$, eliminating a parameter with respect to the generic case. With this choice the relation \eqref{eq:quench152} and higher orders determine $a_{n>0}$ in terms of $b_1/a_0$ and $\phi_0$. Allowing for radial shifts, we are left with four integration constants.
Hence black hole solutions span a co-dimension 1 surface in the space of solutions to EDG with a chosen scalar potential. 

Every $\Om$-family of $f_{1,2}$ functions giving rise to static spacetimes with a (non-extremal) Killing horizon, has an associated smooth solitonic solution with a center. (Similar conclusions hold for stationary solutions.)
This solution is obtained by exchanging the roles of the input functions 
\eq{
f_{1\,s}= f_2 \qquad f_{2\,s}=f_1
}{eq:add09}
and demanding the absence of a conical defect. The absence of a conical defect, condition \eqref{eq:add0}, specifies uniquely the value of $\Omega$ to
\eq{
\Omega_s=\frac{a_0}{16}
}{eq:add009}
where we have used \eqref{eq:add200}.
Therefore, for any $(n+1)$-parameter family of static black holes there is a smooth $n$-parameter family of static solitonic solutions with a center. The simplest example is the 1-parameter family of static BTZ black holes, which has a 0-parameter `family of solitons', namely the global AdS solution. It is suggestive to consider these solitonic solutions as possible ground states of a given theory. 
To decide which geometry is the ground state we have to calculate the free energy, which we shall do in section \ref{se:TD}.

\paragraph{Rotating black holes}
Locally there is no difference between rotating and static black holes in three dimensions. Indeed, all BTZ black holes \cite{Banados:1992wn} --- rotating or static --- are locally AdS and thus locally equal to each other. Rotating black holes differ from static ones by their periodicity properties of the coordinates $x^0$, $x^1$. For the same functions $f_1$ and $f_2$ that produce static black holes with the choices $a=d=1$, $b=c=0$ in \eqref{eq:quench137}, we can pick other values for $a,\,b,\,c,\,d$ to obtain rotating black holes.
In parlance of canonical analysis the $SL(2)$ transformations that we exploited for a local gauge fixing may fail to be gauge transformations globally, since their canonical boundary charges can be non-zero. When studying global solutions we should therefore undo the gauge fixing that led to the Ansatz \eqref{eq:quench11}.
Of the three independent $SL(2)$ transformations, one leads to a rescaling $(\ga f_1,\,f_2/\ga)$, while another one leads to a rotation of the asymptotic metric.
We are not interested here in either of these transformations, but instead want to consider only those $SL(2)$ transformations that keep fixed the asymptotic metric. These transformations are generated by a single parameter $\eta$. 

For black holes in AdS, with our gauge choice $\lim_{\rho\to\infty} f_1/f_2=1$, the transformations are $SO(1,1)$.
\eq{
t = x^0\,\cosh\eta - x^1\,\sinh\eta \qquad \varphi = -x^0\,\sinh\eta + x^1\,\cosh\eta
}{eq:eta}
Static solutions \eqref{eq:quench19} are then the $\eta\to 0$ limit of a larger set of stationary solutions \eqref{eq:ADM} with the identification \eqref{eq:eta}.
Similarly, for asymptotically flat solutions with, say,  $\lim_{\rho\to\infty} f_1=1$, $\lim_{\rho\to\infty} f_2/\rho=2$, the relevant transformations are parabolic, $t=x^0$, $\varphi=-x^0\,\eta + x^1$.\footnote{%
Notice that allowing $b\!\neq\!0$ in this case will induce negative $R^2$ at infinity and thus closed time-like curves. This provides an alternative argument for disregarding one of the three $SL(2)$ parameters in this type of asymptotically flat solutions.
} 
Finally, for dS the transformations are $SO(2)$; but recall that in EDG there are no black holes in dS, as we have shown in the previous section.

It makes sense to consider the parameter $\eta$ as one of the constants of motions, since it parametrizes the black hole angular momentum.
Thus, globally the total number of potentially relevant constants of motions is four: $\{\eta,\,\Omega,\,\xi,\,\lambda\}$.

\paragraph{Concluding remarks}
We have now a clear understanding of the role of nearly all the seven constants of motion: one is fixed trivially by choosing the value of the radial coordinate at the center/singularity, $\rho_0$. Two, related to $SL(2)$ transformations, are fixed by choosing the boundary metric. The remaining four are physically more relevant, with the following meanings. The parameter $\eta$ (related to residual $SL(2)$ transformations) generates rotating solutions from static ones. The parameter $\Omega$ generates a 1-parameter family of solutions for the same scalar potential, in general with different masses. The parameter $\lambda$ will be discussed in more detail in section \ref{se:BF}. 
We finish this discussion speculating on what type of deformation can move away from geometries with a Killing horizon while keeping the same scalar potential, which helps us to identify the role played by the seventh parameter $\xi$.

In the past subsection we have explored the construction of asymptotically flat or (A)dS solutions, taking as input a function $f_1$ positive at large values of the radial coordinate and at least second order differentiable up to its first zero.
The inequality \eqref{eq:quench14} forces $f_1$ to vanish at some finite radial point $\rho_0$. A sufficient condition for the existence of a Killing horizon, which would hide the generically singular behavior at $\rho_0$, is the divergence of the integral $\int^\rho\extd\rho^\prime  f_1^{-2}$ as $\rho$ tends to $\rho_0$. 
Therefore, under the mentioned differentiability assumption of $f_1$, a necessary criterion for the absence of a Killing horizon is the finiteness of the previous integral at $\rho_0$.

The no-Killing-horizon criterion is fulfilled when $f_1\!=\!a_{1-\al} \Om^\alpha (\rho\!-\!\rho_0)^{1-\alpha}$ with $1\!>\!\alpha\!>\!1/2$ close to $\rho_0$. Varying $\alpha$ can be naturally interpreted as a deformation from the case with a horizon if at the same time $f_2\!=\!b_\al \Om^{1\!-\!\alpha}(\rho\!-\!{\rho_0})^\alpha$. A horizon at $\rho_0$ can be obtained when $\alpha$ equals unity, while moving to smaller values of this parameter leads to geometries with a naked singularity.\footnote{Since $f_2$ vanishes at $\rho_0$ faster than $f_1$, expression \eqref{eq:add4} must become zero at $\rho_0$. Given that \eqref{eq:add4} can only have one zero for $\rho\!\geq\!\rho_0$, the associated singularity will be naked.} Moreover, for the proposed $f_1$ and $f_2$ the relation between $\xi$ and $\alpha$ is
\eq{
\xi = 8 \alpha-4
}{eq:add10}  
when, analogously to \eqref{eq:add200}, we fix $a_{1-\al} b_\al=8$ by an appropriate choice of $\Om$.
The complete form of the functions $f_{1,2}$ should be determined by keeping the scalar potential unmodified. Since in the solutions we are considering $\phi$ diverges at $\rho_0$ [see \eqref{eq:add2}], this question concerns the form of the potential at large values of the scalar field. 
The discussion above suggests that varying $\xi$ in general leads to singular solutions.
We analyze next a very simple example where this picture is realized.

\subsection{Solutions with vanishing potential --- gravity plus kinetic energy}\label{se:2.4}

For $V=0$ all solutions can be constructed in closed form (see for instance \cite{Saenz:2012ga}).
This is one of the rare cases where our algorithm can be inverted and taking the potential as input explicitly leads to the functions $f_{1,2}$ as output.
Integrating \eqref{eq:quench8} with the left-hand-side set to zero obtains
\eq{
\vecX^2 = 4\big(\rho^2+c_1\rho+c_2\big)
}{eq:quench31}
where $c_{1,2}$ are integration constants.
Depending on the sign of the discriminant $\De=c_1^2-4c_2$ we obtain different cases.
We discuss explicitly the case of positive discriminant.

With no loss of generality we fix $c_1\!=\!0$ by exploiting shifts and define $c_2\!=\!-\Om^2$.
Integrating the Klein--Gordon equation \eqref{eq:kg} twice yields
\eq{
\phi = \phi_- + \frac{\phi_+}{2\Omega}\, \log {\rho+\Om \over \rho-\Om} 
}{eq:quench32}
where $\phi_\pm$ are integration constants. The value of $\phi_+$ is bounded by $|\phi_+/\Om|\leq 1$. Violating this bound leads to a change of the sign of the discriminant.
Integrating the Einstein equation \eqref{eq:einstein} and imposing the Hamilton constraint \eqref{eq:hamilton} then establishes
\eq{
f_{1,2} = 2  \sqrt{\rho^2-\Om^2} \, \Big(\frac{\rho+\Om}{\rho-\Om}\Big)^{\pm \frac12\,\sqrt{1-\phi_+^2/\Om^2}}
}{eq:quench33}
which introduces one more integration constant. 

The Ricci scalar \eqref{eq:quench25} and the Cotton tensor \eqref{eq:quench26} are given by
\eq{
R=-6 + \frac{2\phi_+^2}{\rho^2-\Om^2} \qquad C_{01} = \frac{4\phi_+^2\sqrt{1-\phi_+^2/\Om^2}}{\rho^2-\Om^2}
}{eq:quench34}
with other components of $C_{\mu \nu}$ vanishing (besides the obvious $C_{10}=C_{01}$). These solutions are always asymptotically AdS. 
If $\phi_+^2\!=\!0$ or $\phi_+^2\!=\!\Om^2$ they are also conformally flat.  
At $\rho^2=\Om^2$ there is a curvature singularity for non-zero $\phi_+$. 

When $\phi_+$ vanishes we obtain vacuum solutions and the function \eqref{eq:quench33} simplifies to
\eq{
f_1=2 (\rho+\Om) \qquad f_2=2 (\rho-\Om)
}{eq:add12}
with a constant scalar $\phi\!=\!\phi_-$. 
The result \eqref{eq:add12} explains why we chose to fix $a_0 b_1=8$ in section \ref{se:2.3}, since we obtain $f_1|_{\rho\approx\Om}=4\Om$ and $f_2|_{\rho\approx\Om}=2(\rho-\Om)$, so $a_0=4$ and $b_1=2$.
This family of solutions should reproduce the BTZ geometries \cite{Banados:1992wn}.
Indeed, redefining the radial coordinate as $r\!=\!\sqrt{2(\rho+\Om)}$ and identifying coordinates as in \eqref{eq:eta} the ADM form \eqref{eq:ADM} of a BTZ black hole is recovered, with mass and angular momentum
\eq{
M = 8\pi\, \Om \cosh2\eta \qquad   J = 8\pi\, \Om \sinh2\eta \,.
}{eq:add14}
Extremal BTZ emerges as a double scaling limit $\Om\to 0$, $|\eta|\to\infty$. An alternative construction of extremal BTZ is possible along the lines of section \ref{se:2.2} and leads to the line-element \eqref{eq:extremalBH}.

When $\phi_+$ is non-vanishing, the behavior around $\rho\!=\!\Om$ is of the form 
\eq{
f_1= 2^{1+\al}\Om^\al\, (\rho-\Om)^{1-\alpha} \qquad f_2 = 2^{2-\al}\Om^{1-\al}\, (\rho-\Om)^\alpha
}{eq:add15} 
with $\alpha\!=\!(1+\sqrt{1-\phi_+^2/\Om^2})/2$; thus $1\!>\!\alpha\!>\!1/2$. We then recover precisely the result \eqref{eq:add10}.
These solutions realize the scheme discussed above. Solutions with a Killing horizon span a co-dimension one subspace in the space of constants of motion. 
The parameter $\xi$ moves away from geometries with a Killing horizon into others with naked singularities. Moreover, the value of $\xi$ is only determined
by the exponent $\alpha$ characterizing the singularity.

The cases of negative and vanishing discriminant lead to complex solutions.
There are no non-trivial real solutions for the metric and the scalar field in these cases.
In conclusion, all non-trivial $V=0$ solutions with real metric and real scalar field have naked singularities, in accordance with the ``no-hair'' property of black holes. 
The only real and regular solutions for vanishing (or constant) potential are vacuum solutions.

\section{Asymptotically flat solutions --- black holes with scalar hair}\label{se:3}

In this section we present an explicit example of locally asymptotically flat black holes with scalar hair.
We define locally asymptotically flat solutions by the property that Ricci scalar and Cotton tensor asymptote to zero for $\rho\to\infty$, and the metric asymptotes to the flat metric in the same limit. The scalar field and its $\rho$-derivative then automatically are finite in the limit of large $\rho$ [see \eqref{eq:quench13}], and the potential goes to a constant [see \eqref{eq:quench8}]. 

Before starting let us briefly summarize a key result from section \ref{se:2}, which leads to the following no-go theorem: there are no globally asymptotically flat hairy black hole solutions in EDG. Indeed, in order to obtain an asymptotically flat spacetime we need that either $f_1$ or $f_2$ tend to a constant at infinity. We saw in \ref{se:2.2} that the existence of a Killing horizon is only possible when the bounded function at infinity is $f_1$. This implies that the space at infinity contains a circle of finite radius, and thus can only be locally asymptotically flat.

Let us therefore consider locally asymptotically flat black holes with scalar hair.
A simple choice for a function $f_1$ approaching a positive constant asymptotically is
\eq{
f_1= 1- \frac{\Om}{\rho}\,.
}{eq:add17}
Reality of the scalar field requires $\Om\!>\!0$. From \eqref{eq:add3} we have
\eq{
f_2= \Om^2 f_1 \Big( A  -2\xi  \Big[ {\Om \over \rho-\Om}-{\rho \over \Om} - 2 \log\big({\rho \over \Om}-1\big)  \Big] \Big)
}{eq:add18}
The choices above with the identification $x^0=t$, $x^1=\varphi$ lead to a line-element that is asymptotically Rindler$_2\times S^1$. 
The function $f_1$ has a simple zero with non-vanishing second derivative. This leads to a curvature singularity at its location $\rho=\Om$
\eq{
R = -6  \xi \log\left({\rho \over \Om}-1\right) + \dots
}{eq:add19}
The singularity is always covered by a Killing horizon, since the term in parenthesis in \eqref{eq:add18} has one and only one zero for any $A$, $\Om$ and $\xi$.
The scalar field is obtained integrating \eqref{eq:quench13}.
\eq{
\phi = \phi_0 + \sqrt{2}\pi -  2\sqrt{2}\arctan\sqrt{\frac{\rho}{\Om}-1}
}{eq:quench52}
It has a divergent kinetic energy at the singularity, but it is regular otherwise. Thus, these solutions are explicit examples of asymptotically flat black hole solutions with non-trivial scalar hair.

The associated scalar potential is derived as a function of the radial coordinate from \eqref{eq:quench15}. The inversion $\rho\!=\!\rho(\phi)$ is easy in this case, and we obtain
\eq{
V=2 + \xi \,\sin^4\hat\phi  + 
    \big(A -  2\xi +  4 \xi \ln{\cot^2\hat\phi}\big) \sin^6\hat\phi  
  - \tfrac32\,  \big(A + 4 \xi + 4 \xi \ln{\cot^2\hat\phi}\big)\sin^8\hat\phi 
}{eq:quench53}
where we have defined $\hat\phi\!=\!(\phi\!-\!\phi_0)/\sqrt{8}$. As expected the potential is independent of the parameter $\Om$. It does, however, depend on $A$ and $\xi$. In subsection 2.3 we have considered $\xi$ as one of the constants of motion that parametrize solutions of EDG with the {\it same} potential. Preserving $V$ required a very non-trivial dependence of both $f_{1,2}$ on $\xi$, codified in \eqref{eq:add6}, \eqref{eq:add7} and \eqref{eq:quench152}. Since our algorithm does not tell us how to solve those equations, we have constructed the present example based on a $\xi$-independent function $f_1$. When $\xi$ is implemented as constant of motion, we argued that it generically moves away from solutions with a Killing horizon into ones with naked singularities. This is avoided in the present case by the explicit dependence of the potential on this parameter. 

\begin{figure}
 \begin{center}
 \epsfig{file=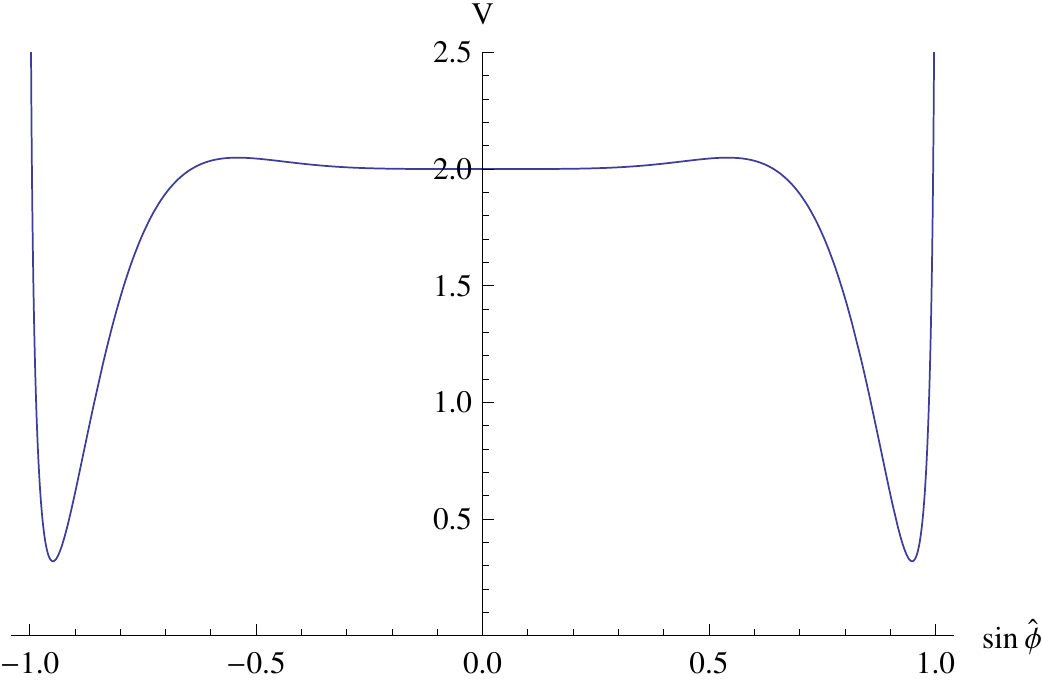}
 \end{center}
\caption{Potential for the exact solution 
(3.5) plotted as function of $\sin\hat\phi$ for the choices $\Om=1$, $A=2$, $\xi=\tfrac12$ and $\phi_0=0$}
\label{fig:1}
\end{figure}

The potential \eqref{eq:quench53} is plotted in figure \ref{fig:1}. In addition to $A$ and $\xi$, it depends on the integration constant of our algorithm $\phi_0$ through the redefined field $\hat\phi$. It has a local minimum at $\phi\!=\!\phi_0$, two global minima lying symmetrically around $\phi\!=\!\phi_0$ and diverges if $\sin\hat\phi\!\to\!\pm 1$. 
It is important to recall that only for a constant potential the quantity $\phi_0$ is a constant of motion in the sense considered above. 

To the best of our knowledge solutions \eqref{eq:add17}-\eqref{eq:quench53} are the first explicit examples of locally asymptotically flat black hole solutions with non-trivial scalar hair in 3-dimensional EDG.
%
The evasion of no-hair theorems has been possible, because these theorems assume typically $V\geq 2$ (see \cite{Sudarsky:1995zg} 
and Refs.~therein), which is easy to violate.
As evident from figure \ref{fig:1} the potential \eqref{eq:quench53} violates the inequality $V\geq 2$ in some range of the coordinate $\rho$. 
Physically the evasion of no-hair arguments is possible due to the presence of a minimum below the asymptotic value of the potential, $\lim_{\rho\to\infty} V=2$; see e.g.~the discussion in \cite{Park:2008zzb}.


\section{Asymptotically AdS solutions and scalar asymptotics}\label{se:BF}

From the line-element \eqref{eq:quench18} we see that in order to obtain an asymptotically AdS metric with
a Minkowski boundary metric the functions $f_{1,2}$ must tend to the same linear function at large $\rho$, namely 
\eq{
f_{1,2}= 2 \rho +\cdots
}{eq:add20}
where the dots stand for subleading terms and the overall factor is fixed by requiring unit AdS radius.  
This fixes the boundary metric and so it corresponds to a Dirichlet boundary condition for the metric. However, in order to properly 
define the variational problem for EDG we need, in addition, to prescribe suitable boundary conditions for the scalar field $\f$. In the current 
section we discuss a simple class of boundary conditions for the metric and scalar fields that preserve  the AdS asymptotics and which  
suitably generalize the Brown--Henneaux boundary conditions for pure gravity \cite{Brown:1986nw}. A complete analysis of the most general 
boundary conditions for EDG gravity that lead to AdS asymptotics is performed in section \ref{se:4}.

\subsection{Boundary conditions --- Brown--Henneaux and beyond}

The Brown--Henneaux boundary conditions for vacuum 3D Einstein gravity with a negative cosmological constant correspond to the Dirichlet
boundary condition \eqref{eq:add20} for the metric. However, this leading asymptotic behavior in fact determines through the EOM
certain subleading terms as well. In particular, for vacuum Einstein gravity, the Brown--Henneaux boundary conditions 
are\footnote{The notation $F(x)=o(f(x))$ as $x\to x_0$ means $\lim_{x\to x_0} \frac{F(x)}{f(x)}=0$.} 
\eq{
f_{1,2} = 2\, \rho + C_{1,2} + o(1)
}{eq:quench59}
where $C_{1,2}$ are arbitrary constants, whose difference determines the parameter $j$ \eqref{eq:quench12}. Introducing a canonical radial coordinate through
\beq
r=\exp \int \frac{d\r}{\sqrt{f_1(\r)f_2(\r)}}\sim c\sqrt{\r}\left(1+\frac{C_1+C_2}{8\r}+\cdots\right)
\eeq 
where $c\neq 0$ is an arbitrary integration constant, 
the line-element \eqref{eq:quench18} becomes
\eq{
\extd s^2 = {\extd r^2 \over r^2} + \Big(g^{(0)}_{ij}\,r^2 + g^{(2)}_{ij} + g^{\rm rest}_{ij}\Big)\,\extd x^i\extd x^j\,
}{eq:quench67}
where
\eq{
g^{(0)}=2c^{-2}\textrm{diag}(-1,\,1) \qquad g^{(2)}={C_1\!-\!C_2 \over 2}\,\textrm{diag}(1,1) \qquad g^{\rm rest}_{ij}=o(1)
}{eq:add22a}
The expansion \eqref{eq:quench67} contains precisely the leading background term and the subleading state-dependent 
term $g^{(2)}$ that appear in the original Brown--Henneaux analysis \cite{Brown:1986nw}. However, neither $g^{(2)}$, nor $g^{\rm rest}$ 
are part of the specification of the boundary conditions. In the case of vacuum Einstein equations one in fact finds that $g^{\rm rest}=\co(r^{-2})$, but 
this will not in general be the case for EDG. 

For EDG instead, Brown--Henneaux boundary conditions correspond to the asymptotic expansions  
\eq{
f_{1,2} = 2\, \rho + C_{1,2} + \co(\r^{-\ve})
}{eq:quench591}
for some unspecified $\ve>0$. Inserting these into the equation \eqref{eq:quench13} for the scalar field leads to the corresponding asymptotic form of the scalar. Namely,
\beq\label{BH-exp-scalar}
\f(\r)=\co(\r^{-\frac{1+\ve}{2}})
\eeq
This asymptotic behavior in turn determines 
the mass term in the expansion of the potential around $\phi\!=\!0$ 
\eq{
V = \frac12 m ^2\,\phi^2 + \cdots
}{eq:quench72}
where the scalar mass is determined in terms of $\ve$ as 
\eq{
m^2 =-(1-\ve^2)
}{eq:angelinajolie} 
It follows that EDG can admit asymptotically AdS solutions provided the parameter $\ve$ introduced in \eqref{eq:quench591} lies in the range 
\beq
0\leq \ve \leq 1
\label{eq:eps}
\eeq
where the lower limit corresponds to the BF bound $m^2_{\textrm{\tiny BF}}=-1$ and the upper one to $m^2=0$. However, once the scalar mass 
is determined so are the two possible asymptotic behaviors of the scalar field, namely
\beq \label{BH-exp-scalar2}
\f(\r)\sim \co(\r^{-\frac{\D_\pm}{2}})
\eeq
with
\eq{
\De_\pm = 1 \pm \eps\,. 
}{eq:quench111}
$\D_+$ then corresponds to the Brown--Henneaux boundary conditions \eqref{BH-exp-scalar}, while $\D_-$ corresponds to a more general class of boundary conditions. 

Since the scalar field is determined from the input function $f_1(\r)$ via the ODE \eqref{eq:quench13}, in order to generalize the boundary conditions and capture both possible asymptotic behaviors of the scalar we need to suitably generalize the asymptotic expansions \eqref{eq:quench591}. This can be done systematically and in complete generality. We do this in section \ref{se:4}. For now, we discuss the simplest possible generalization of \eqref{eq:quench591}, which takes the form
 \beq\label{eq:quench60}
 f_1 = 2\rho + A\Om^{1-\eps}\,\rho^\eps + B\,\Om\,\log\rho + C \, \Om + \cdots
 \eeq
where $A$, $B$, $C$ and $\Om\neq 0$ are constants, and the dots stand for terms suppressed at large $\rho$. Moreover, we assume $0<\ve <\frac12$ for now.

Given the asymptotic expansion for the input function $f_1(\r)$, the constraint \eqref{eq:quench12} can be integrated to obtain 
\beq
f_2 = f_1 - \xi\, \Om + \cdots  
\eeq
with $j=\xi \, \Om$.
Moreover, inserting the expansion \eqref{eq:quench60} into \eqref{eq:quench13} we get
\eq{
\dot\phi = -\Om^{(1-\eps)/2}\,\sqrt{\frac{\eps(1-\eps)A}{2}} \,\rho^{(\eps-3)/2} - \frac{B\, \Om^{(1+\eps)/2}}{4}\,\sqrt{\frac{2}{A\eps(1-\eps)}}\,\rho^{-(\eps+3)/2} + \cdots
}{eq:quench70}
which can be integrated to obtain
\eq{
\phi = \phi_-\,\rho^{-\De_-/2} + \phi_+\,\rho^{-\De_+/2} + \cdots
}{eq:quench71}
where
\eq{
\phi_-=\a\;\Om^{\De_-/2} \qquad \phi_+=\frac{B}{\a\D_-\D_+} \Om^{\De_+/2}
}{eq:quench117}
and we have defined 
\beq
\a\equiv\sqrt{\frac{2 A\ve}{1-\ve}}\,. 
\eeq
We see that the input function \eqref{eq:quench60} leads to the correct asymptotic form of the scalar with both possible asymptotic modes.
It is therefore a valid generalization of the Brown--Henneaux boundary conditions for $0<\ve <1/2$.  

The parameter $B$ is proportional to the integration constant $\lambda$ defined in  \eqref{eq:add5} 
\eq{
B= \De_- \De_+  \alpha^{2 \over \De_-} \, \lambda 
}{eq:add202}
The proportionality constant is set by the exponent $\eps$, which fixes   
the scalar mass and thus is a characteristic of the potential. We have    
argued in subsection \ref{se:2.3} that starting with a regular solution   
and varying $\xi$ while keeping $\Om$, $\lambda$ and the scalar potential
fixed, in general leads to singular solutions. 
Since mapping to the dual CFT is only
established for regular solutions, we will not consider the dependence of 
solutions on $\xi$. On the other hand, implementing $B$ as integration
constant would require a very non-trivial dependence on this parameter of
the $f_1$ piece unspecified in \eqref{eq:quench60}, on which we will not 
enter. As extensively discussed only the $\Om$-dependence of solutions is
immediate to implement, which we have done in \eqref{eq:quench60}.

The usual AdS/CFT dictionary interprets $\phi_-$ as a source term in the dual CFT. According to this, the 1-parameter family of solutions generated by $\Om$ would move among boundary field theories with different couplings. 
However, since both modes of a scalar with mass in the window \eqref{eq:angelinajolie}, \eqref{eq:eps} are normalizable, we can choose not only $\phi_-$ as a source
(Dirichlet boundary conditions), but also $\f_+$ (Neumann boundary condition) or any combination of the modes $\f_-$ and $\f_+$ \cite{Witten:2001ua}. Especially interesting boundary conditions are those that allow to vary $\Om$ without modifying the boundary field theory. Clearly this is achieved if we impose 
\eq{
\phi_+ - \lambda \phi_-^{\Delta_+/ \Delta_-}=0
}{eq:add201}
with $\lambda$ fixed. This boundary condition corresponds \cite{Witten:2001ua} to subject the CFT associated to Neumann boundary conditions to a marginal multitrace deformation
\eq{
\De L=\lambda\, {\cal O}_{\De_-}^{2 \over \De_-}
}{eq:add23}
where ${\cal O}_{\De_-}$ is the single trace operator dual to $\f_+$. 
Black hole families generated by $\Om$ provide then the holographic dual for certain thermal states in such a deformed CFT. In section 6 we will discuss their thermodynamic stability, as well as the role of the associated soliton geometry \eqref{eq:add09}, \eqref{eq:add009}  \cite{Correa:2010hf,Correa:2011dt}. 

Since we want to study the thermodynamics of the dual field theories, we need to address the holographic renormalization of our backgrounds. Although practical reasons highlight marginal boundary conditions, any other ones are a priori of equal relevance. Notice that the stability of a solution depends crucially on the chosen boundary conditions.
We will start below the analysis of holographic renormalization considering Dirichlet boundary conditions.
Neumann and Mixed boundary conditions will be treated in section \ref{se:gbc}.



\subsection{Holographic renormalization}\label{se:hol}

The bulk action for field configurations solving the EOM can be rendered finite by the addition of appropriate boundary counterterms  
\eq{
\Gamma_+ =\int\extd^3x\sqrt{-g}\,\Big[R+2-\frac12\,(\partial\phi)^2-V(\phi)\Big] + 2\int_\partial \extd^2x\sqrt{-\ga}\,\Big[K-U(\phi)\Big]
}{eq:quench17}
where the subscript `$+$' indicates that we choose Dirichlet boundary conditions for the scalar. $K$ is the trace of the extrinsic curvature on the boundary $\partial$, 
defined as a $\rho\!=\!\rm const.$ surface for asymptotically large values of the radial coordinate, and $\ga$ is the induced metric on $\partial$. 
For solutions corresponding to an input function of the form \eqref{eq:quench60} 
the counterterm function $U(\f)$ is given by
\eq{
U(\phi) = 1 + \frac{\De_-}{4}\,\phi^2. 
}{eq:quench118}
We will derive this result in section \ref{se:4}, where the form of $U(\f)$ for the most general asymptotically AdS solutions 
is also derived. However, we emphasize the results we give in the following for the renormalized action and one-point functions are in fact valid for the most general 
asymptotically AdS solutions that depend only on the radial coordinate, as can be explicitly confirmed using the counteterms derived in section \ref{se:4}. Moreover,  
it should be stressed that the counterterms in \eqref{eq:quench17} are not sufficient for regularizing general solutions, but only those which solely depend 
on the radial coordinate. Since in this paper we are considering stationary, axisymmetric solutions of EDG, it is enough for our purposes. 

Taking into account all boundary terms, the on-shell action is
\begin{eqnarray}
\label{eq:quench16}
\Gamma_+\big|_{\textrm{EOM}} & = & 
{\textrm{Vol} \over 2} \left( \partial_\rho(f_1 f_2) |_{\rho_h}+ \lim_{\rho\to\infty} \Big[  \partial_\rho (f_1 f_2) -4 U(\phi)\sqrt{f_1 f_2}  \Big] \right)\\
& = & {\rm Vol}\,\Big( |j|+\De_- (\De_+\!-\!\De_-)\f_-\f_+\Big) 
\end{eqnarray}
with $\textrm{Vol}\!\equiv\!\int \!\extd x^0 \! \extd x^1$.
The quantity $\rho_h$ denotes the locus of a Killing horizon or center, which is the only natural lower bound of the radial integration domain. 
The discussion in section \ref{se:2.3} shows that when the asymptotic signs of $f_1$ and $f_2$ coincide  the reduced model angular momentum $j$ is positive for metrics with a Killing horizon while negative for those with a center. Hence the first term on the right hand side of \eqref{eq:quench16} defines the absolute value of $j$ on solutions of the form \eqref{eq:quench60}.

The variation of the full on-shell action \eqref{eq:quench17} with respect to the induced metric
\eq{
\de\Gamma_+\big|_{\textrm{EOM}} = \int_\partial \extd^2x\sqrt{-\ga}\,\Big[K_{ij}-(K-U(\phi))\ga_{ij}\Big]\,\de\ga^{ij}
}{eq:quench78} 
determines the holographically renormalized Brown--York stress tensor, i.e.~the renormalized one-point function of the stress tensor of the dual field theory, namely 
\eq{
\langle T_{ij} \rangle_+ = -{2 \over \sqrt{-g^{(0)}}}  \left.{\de\Ga_+ \over \de g^{(0)ij}}\right|_{g^{(0)}=2\h}
}{eq:quench81}
where we have set the constant $c=1$ in \eqref{eq:add22a} for solutions of the form 
\eqref{eq:quench60} so that $\ga_{ij}\!=\!g_{ij}^{(0)} \rho+\dots$, with $g_{ij}^{(0)}\!=\!2 \eta_{ij}$. Evaluating this we obtain 
\eq{
\langle T_{ij}\rangle_+= j \, \delta_{ij} + \De_- (\De_+\!-\!\De_-)\f_+\f_- \, \eta_{ij} 
}{stresstensor}
whose trace is
\eq{
\langle T^i{}_i\rangle_+ = g^{(0)ij} \langle T_{ij}\rangle_+
= \De_- (\De_+\!-\!\De_-)\phi_+ \phi_-
}{eq:quench119}
Notice that if either $\phi_\pm$ vanish then the stress tensor is traceless. If additionally the functions $f_i$ are proportional to each other then $j$ vanishes, the spacetime is conformally flat and the energy momentum tensor vanishes completely.

Another interesting object is the response with respect to variations of the scalar field, which provides the expectation value of the operator dual to the $\f_-$, $\f_+$ combination that we chose as source. Dirichlet boundary conditions yield
\eq{
\de\Gamma_+\big|_{\textrm{EOM}} = -\int_\partial \extd^2x\sqrt{-\ga}\,\Big[n^\mu\partial_\mu\phi+2U'(\phi)\Big]\, \de\phi
}{eq:quench100} 
where $n^\mu$ is the outward pointing unit normal vector to the boundary $\partial$. The variation of the scalar field associated with the mode $\phi_+$ gives an asymptotically vanishing contribution to \eqref{eq:quench100}, and therefore only $\de \phi\!=\! \de \phi_- \, \rho^{-\De_-/2}$ is relevant. 
The vacuum expectation value of the field theory operator dual to $\phi_-$ then reads
\eq{
\langle{\cal O}_{\De_+}\rangle = -{1 \over \sqrt{-g^{(0)}}}  {\de\Ga_+ \over \de \phi_-} 
= -(\De_+\!-\!\De_-) \phi_+ 
}{eq:quench116}
The ratio between the vacuum expectation value of the trace of the stress tensor and the scalar response is a $\phi_+$-independent constant
\eq{
\langle T^i{}_i\rangle_+ 
= -\De_- \phi_- \, \langle{\cal O}_{\De_+}\rangle 
}{eq:quench83}
This is the usual trace Ward identity. 

\subsection{Beyond Dirichlet boundary conditions}\label{se:gbc}

We will analyze now Neumann and Mixed boundary conditions. We denote as $\cj(\f_-,\f_+)$ the combination of the two scalar modes that one decides to take as source.
We can have 
\begin{subequations}
\label{general-bc}
\begin{align}
&&& {\rm Dirichlet} &  \cj &:=\f_-  \\
&&& {\rm Neumann} &  \cj &:=
(\De_+\!-\!\De_-) \f_+  \\
&&& {\rm Mixed} &  \cj &:=
(\De_+\!-\!\De_-) \f_+-\cw'(\f_-) 
\end{align}
\end{subequations}
where $\cw(\f_-)$ is an arbitrary function. The field theory operator sourced by $\cj$ has conformal dimension $\De_+$
in the case of Dirichlet boundary conditions and $\De_-$ in the cases of Neumann or Mixed boundary conditions.  
 
In order to have a well posed variational principle, the change of the action under variations of the fields must be proportional to the variation of the sources (modulo the EOM). Indeed the sources are naturally to be held fixed for a dual field theory interpretation.
This justifies the claim that $\Ga_+$, as defined in \eqref{eq:quench17}, corresponds to Dirichlet boundary conditions. Namely, its variation
is proportional to $\d\f_-$. Similarly, it follows that in order to impose Neumann or Mixed boundary
conditions the following boundary term must be added to \eqref{eq:quench17}:
\beq\label{extra-boundary-term}
S_\cw=-\int \extd^2x \sqrt{-g^{(0)}}  \Big[ (\De_+\!-\!\De_-)\f_+\f_- + \cw(\f_-)-\f_-\cw'(\f_-) \Big]
\eeq
The Neumann case is obtained by setting $\cw=0$. Adding this extra boundary term we find 
that the change of the on-shell action under variations of the scalar field is
\beq
\de\Ga_- = \d(\Ga_+ + S_\cw)=- \int \extd^2x \sqrt{-g^{(0)}}  \, \f_-\d \cj 
\eeq 
which is proportional to $\d\cj$ as required. An immediate conclusion we draw from this identity is that the scalar one-point function 
for Neumann or Mixed boundary conditions is given by
\beq
\langle \co_{\D_-} \rangle = - {1 \over \sqrt{-g^{(0)}}} \,\frac{\d \Ga_-}{\d\cj\;}=\f_- 
\label{eq:add24}
\eeq 

The one-point function of the stress tensor for Neumann or Mixed boundary conditions is given 
by \cite{Papadimitriou:2007sj} 
\beq\label{eq:stress-tensor-vev}
\langle T_{ij}\rangle _-=\langle T_{ij}\rangle _+-\left(\cw(\f_-)+\f_-\cj\right) g^{(0)}_{ij}
\eeq
The trace of the stress tensor yields
\eq{
\langle T^i{}_{i}\rangle _-
=-\De_+\f_-\,\cj + \De_- \f_- \,\cw'(\f_-) - 2 \cw(\f_-) 
}{eq:lalapetz}
Notice that for marginal boundary conditions \eqref{eq:add201}, \eqref{eq:add23} we have
\beq\label{marginal-bc}
\cw(\f_-)=\l {\De_-(\De_+ \!-\! \De_-) \over 2}\; \f_-^{\frac{2}{\De_-}}
\eeq
and the trace Ward identity simplifies to 
\beq
\langle T^i{}_{i}\rangle _- = -\De_+ \cj\langle \co_{\D_-} \rangle\,. 
\label{eq:add104}
\eeq
This is exactly of the form \eqref{eq:quench83}, except that expectation values and sources have been 
changed and the dual scalar operator in this case has dimension $\D_-$.  

We conclude this section with a couple of observations. Rewriting the boundary term \eqref{extra-boundary-term} as 
\beq
S_\cw= -\int \extd^2x \sqrt{-g^{(0)}}  \Big[ \f_- \cj + \cw(\f_-) \Big]
\eeq
and using the second equality in \eqref{eq:add24},
it follows that the effective action for the scalar expectation value $\f_-$, given by the Legendre transform, takes the form
\beq
\G[\f_-] = \Ga_- - \int \extd^2x  \; \cj \, \frac{\d \Ga_-}{\d\cj}= \Ga_+ -\int \extd^2x  \sqrt{-g^{(0)}}\;\cw(\f_-)\,.
\label{eq:add25}
\eeq
Hence the function $\cw(\f_-)$ corresponds to a multitrace deformation of the theory corresponding to 
pure Neumann boundary conditions. 

A second remark concerns the relation between boundary conditions and regularity of the solutions. 
States in the theory, i.e.~regular solutions, corresponding to Mixed boundary conditions satisfy 
\beq
\cj=(\D_+-\D_-)\f_+-\cw'(\f_-)=0.
\eeq
This imposes a relation between the two modes $\f_-$ and $\f_+$. However, the condition that the solutions be also regular imposes {\em another} condition, 
which can be written in the form 
\beq\label{mixed-ct}
(\D_+-\D_-)\f_+-\partial_{\phi_-}\cw_{\textrm{reg}}(j,\,\f_-)=0,
\eeq
for an a priori different function $\cw_{\rm reg}(j,\, \f_-)$. Combining this with \eqref{eq:quench116} and \eqref{eq:add25} we get 
\beq
\G[\f_-] = \int \extd^2x  \sqrt{-g^{(0)}}\left[\cw_{\rm reg}(j,\,\f_-)-\cw(\f_-)\right]
\label{effective-potential}
\eeq
up to an unphysical constant. This is the full effective potential for the VEV of the dual operator $\co_{\D_-}$. Smooth EDG solutions correspond to extrema of this effective potential, i.e.~they must satisfy $\partial_{\phi_-}(\cw_{\rm reg}-\cw)=0$.  

In section \ref{se:2.3} we discussed the criterium for regularity of solutions in term  of the constants of motion $\{\Om,\, \xi,\, \lambda\}$. We showed that $\xi$ is fixed by regularity.
Moreover, defining $\Om$ to satisfy \eqref{eq:add200} lead to a constant value of $\xi$ on regular solutions. This convenient choice was possible due to the freedom to redefine $\Om$ by a multiplicative factor that can depend on the other constants of motion, which is manifest in \eqref{eq:quench147}. In this section we have performed an analysis of solutions of EDG based only on their asymptotic properties, while imposing \eqref{eq:add200} requires the knowledge of the interior geometry. Hence for the parameterization \eqref{eq:quench60}, the regularity condition will instead read
\eq{
\xi=\xi(\lambda)
}{eq:add250}
This relation is equivalent to \eqref{mixed-ct}. 

Once the regularity condition is imposed, the smooth solutions are described by the parameters $\{\Om,\, \lambda\}$, or equivalently $\{j,\, \phi_-\}$ (and globally in addition by the rotation parameter $\eta$). If we impose marginal boundary conditions \eqref{marginal-bc} the parameter $\lambda$ is fixed for all solutions compatible with these boundary conditions. For other boundary conditions $\lambda$ varies on the solutions.

\section{Asymptotically AdS solutions --- general discussion}\label{se:4}

In this section we generalize the discussion of the previous section by extending the range of allowed values of
$\eps$ in the asymptotic expansion \eqref{eq:quench60} to the whole interval $0\leq \eps \leq 1$. In particular, 
we determine the most general form of the input function $f_1(\r)$ such that the theory admits AdS solutions and 
can be holographically renormalized by local counterterms. On the way we determine the general form of 
the counterterms in \eqref{eq:quench17}, as well as the most general asymptotic form of the scalar potential.

\subsection{Most general asymptotically AdS solutions}\label{se:mostgen}

It must be emphasized  that we are not interested in any possible theory, i.e.~scalar potential, that admits asymptotically AdS solutions. 
Instead we are interested in are theories that not only 
admit asymptotically AdS solutions, but {\em also} can be holographically renormalized with {\em local}
covariant boundary terms. The latter is a necessary condition for the locality of the holographically
dual CFT and constrains the form of the scalar potential. From the point of view of the algorithm 
presented in this paper, the requirement that the dual CFT can be holographically renormalized 
restricts the asymptotic form of the input function $f_1(\r)$. 

The asymptotic expansions are qualitatively different for the cases $\eps=0$, $\eps=1$ and $0<\eps<1$. 
Starting with the latter case one finds that the most general asymptotic form of the input function that appropriately generalizes the expansion 
\eqref{eq:quench60} for generic $ 0 < \eps < 1$ is 
\beq\label{eq:mostgen}
 f_1 = 2\r+ A_2\,\Om^{1-\ve}\,\r^{\ve}+ \sum\limits_{m=3}^n\,A_m\,\Om^{\frac{m(1-\ve)}{2}}\,\r^{1-\frac{m(1-\ve)}{2}}
+\tilde B \Om\,(\log\r)^2+ B\,\Om\,\log\r+C\Om + \cdots\\
\eeq
where the constants $\ve$, $A_k$, $\tilde B$, $B$, $\Om$ and $C$ are the parameters specifying the input function.
It should be stressed that even though these parameters are the input and can in principle be prescribed at will, this in fact does {\em not} always lead to 
a scalar potential of the desired form. In general there are some constraints that these input parameters must 
satisfy, which we will discuss momentarily. For $0<\ve <1$ the parameter $\ve$ defines a unique integer $n\geq 2$ via the inequalities 
\beq\label{condition}
\frac{2}{n+1}\leq 1-\ve <\frac 2n\,,
\eeq
which tell us how many terms have to be included in the asymptotic expansion \eqref{eq:mostgen}. The
infinite set of special values of the parameter $\ve$, corresponding to the case when the first inequality in  (\ref{condition}) is saturated, defines the series of so called resonant scalars 
(see e.g.~\cite{Banados:2006de} and references therein). 
\eq{
\textrm{resonant\;scalars:}\qquad\qquad\frac{\De_+}{\De_-} = n
}{eq:resonant}
For $0<\eps<\tfrac13$ (or $0<\eps<\tfrac12$, 
together with assuming vanishing $A_3$) the sum in the first line of \eqref{eq:mostgen} disappears and we 
are back to the case discussed in the previous section. 

Given the asymptotic expansion for the input function $f_1(\r)$ we can obtain the asymptotic form of 
the function $f_2(\r)$ by integrating (\ref{eq:quench12}) and that of the scalar $\f(\r)$ 
by integrating (\ref{eq:quench13}). For $f_2$ we get
\beq
 f_2 = f_1 - \Om\,\xi + \cdots 
\eeq
where $j=\Om\,\xi$ is again the angular momentum defined in \eqref{eq:quench12}. For the scalar field one finds
\bea\label{scalargeneral}
\f(\r) &=&\r^{\frac{\ve-1}{2}}\left(\f_-+\m_3\f_-^2\r^{\frac{\ve-1}{2}}+\cdots+
\m_n\f_-^{n-1}\r^{\frac{(n-2)(\ve-1)}{2}}+\tilde\m_{n+1}\f_-^{n}\r^{-\ve}\log\r\right.\NO\\
&&\left.\phantom{evenmoremoremoremorespacehere}+\f_+\r^{-\ve}+\cdots\rule{0.0cm}{0.389cm}\right),
\eea
where the coefficients $\phi_-$, $\m_k$, $\tilde \m_{n+1}$ and $\f_+$ are determined in terms of the parameters of the input function as
\begin{align}
&\phi_-=\a\;\Om^{\De_-/2} \qquad \mu_3=\frac{3(3\ve-1)}{8(1-\ve)}a_3\qquad
\mu_4 = -\frac{2  (1-2\eps)}{3(1-\eps)}a_4-\frac{1-\eps}{24\eps}-\frac{2}{3}\mu_3^2\quad \cdots\NO\\
&\tilde\m_{n+1}=\frac{2\tilde B}{\a\D_-\D_+} \Om^{\De_+/2}
\qquad \phi_+=\frac{1}{\a\D_-\D_+} \Big(B-\frac{2\tilde B}{\D_+}\Big)\Om^{\De_+/2}  \label{B}
\end{align}
with 
\beq
\a\equiv\sqrt{\frac{2 A_2\ve}{1-\ve}} 
\qquad a_k\equiv \frac{A_k}{\f_-^k}\Om^{\frac{k\D_-}{2}}=\frac{A_k}{\a^k}
\eeq
introduced for convenience. It is clear from these expressions that, as alluded to above, the parameters specifying
the input function $f_1(\r)$ cannot in fact be chosen completely at will. In particular, the inequality \eqref{eq:quench14} requires $A_2\geq 0$.\footnote{%
The case $A_2=0$ is special since the asymptotic expansion
for the scalar is apparently ill defined in this case. However, one should keep in mind that the above expressions
for the coefficients of the asymptotic expansions are in general expressions for $A_k$, $\tilde B$ and $B$ as 
functions of the parameter $\phi_-$ and they remain valid even for $\phi_-=0$. For nonzero $\phi_-$, or $A_2$, we
can invert these relations and treat $A_k$, $\tilde B$ and $B$ as input, but this is not possible when $A_2=0$. 
We conclude that if $A_2=0$, then all $A_{k>2}$, $\tilde B$ and $B$ must be zero in order to obtain 
a scalar potential that admits an AdS vacuum, while $\phi_+$ is arbitrary. This case implies Brown-Henneaux boundary conditions $\f_-=0$.
} 
Yet another restriction concerns the parameter $\tilde B$. This 
parameter can be nonzero only in the case of resonant scalars since in all other cases it leads to a 
non-renormalizable theory.

For $\ve=0$ the mass \eqref{eq:angelinajolie} saturates the BF bound and the asymptotic expansion of 
the input function $f_1(\r)$ takes the form
\beq
 f_1 = 2\r+ A\Om (\log\r)^3+\tilde B \Om(\log\r)^2+ B\,\Om\,\log\r+C\Om + \cdots\\
\eeq
Integrating (\ref{eq:quench13}) in this case leads to 
\beq\label{BF-expansion}
\f(\r)=\r^{-1/2}\left(\log\r\; \f_-+\f_++\cdots\right),
\eeq
where
\beq
\f_-=\sqrt{6A}\Om^{\frac12},\qquad \f_+=\sqrt{\frac{2}{3A}}\left(3A+\tilde B\right)\Om^{\frac12}\,. 
\eeq

Finally, for $\ve=1$ (vanishing mass) the input function $f_1(\r)$ must be of the form
\beq
 f_1 = 2\r+C\Om+\frac{D\Om^2}{\r}+ \cdots\\
\eeq
leading to 
\beq
\f=\f_-+\f_+\r^{-1}+\cdots
\eeq
where $\f_+=\sqrt{-D}\Om$ and $\f_-$ is the arbitrary integration constant of (\ref{eq:quench13}). 

\subsection{Scalar potential}\label{se:potential}

The asymptotic expansions of the previous subsection were constructed by demanding that the corresponding 
scalar potential $V$ takes a certain asymptotic form so that the theory can be holographically renormalized. 
In this subsection we determine the potential $V$ that follows from the asymptotic expansions above.

Again, one has to distinguish the cases $\eps=0$, $\eps=1$ and $0<\eps<1$. In the latter case, the most 
general scalar potential that follows from (\ref{eq:mostgen}) takes the 
form 
\bea\label{general-ads-potential}
V(\phi)=
\frac12 m^2\f^2+v_3\f^3+\cdots+v_n\f^n+v_{n+1}\f^{n+1}+o(\f^{n+1})
\eea
where the mass is given by \eqref{eq:angelinajolie} and the next couple of terms are
\beq
v_3=-\frac18(3\ve-1)^2a_3\quad
v_4=-\frac23(2\ve-1)^2a_4+\frac{3(3\ve-1)^2(17\ve-7)}{128(1-\ve)}a_3^2-\frac{(7\ve-2)(1-\ve)^2}{24\ve}
\label{eq:nolabel2}
\eeq
Note that the cubic coefficient $v_3$ is non-vanishing only if the expansion coefficient $A_3$ is non-zero. By contrast, the quartic coefficient $v_4$ is in 
general non-zero even when both $A_3$ and $A_4$ vanish. 
Moreover, at order $n+1$ there is a difference depending on whether the inequality \eqref{condition} is saturated or not. 
Namely, in the case of resonant scalars the coefficient $v_{n+1}$ contains an additional contribution relative to
the non-resonant case, given by
\beq
-2\ve(1-\ve)\tilde\m_{n+1}.
\eeq

By contrast, in the case $\ve=0$ the scalar potential is much less constrained and takes the generic form
\beq\label{BF-potential}
V(\f)=-\frac12\f^2+o(\f^2)
\eeq 
while for $\ve=1$ the potential vanishes, leading to the case discussed in section \ref{se:2.4}.

\subsection{Holographic renormalization}\label{se:hr}

Now that we specified the scalar potentials we are interested in, we determine systematically the 
counterterm function $U(\f)$ in \eqref{eq:quench17} in this subsection.
We can do this directly at the level of the minisuperspace model \eqref{eq:quench5}
using a Hamiltonian language where the radial coordinate plays the role of Hamiltonian `time'. 
In the full Einstein-scalar theory this amounts to an ADM \cite{Arnowitt:1960es} formulation of the dynamics in 
the radial coordinate.

From \eqref{eq:quench5} (with $S=\int\extd\rho\, L$) we obtain the canonical momenta
\beq\label{momenta}
{\Pi}_i=\frac{\pa L}{\pa\dot X^i}={\rm Vol}\;e^{-1}\h_{ij}\dot X^j \qquad \P_\f=\frac{\pa L}{\pa\dot\f}=-{\rm Vol}\;\vecX^2\dot\f\,.
\eeq
At the same time, these momenta are related to the {\em on-shell} action, or Hamilton's principal function
$\cs[X,\f]$, by  
\beq\label{HJ-momenta}
\P_i=\frac{\d \cs}{\d X^i} \qquad \P_\f=\frac{\d\cs}{\d \f}\,.
\eeq
Combining these two expressions for the canonical momenta allows us to write the constraints \eqref{eq:hamilton} 
and \eqref{eq:quench7} as PDEs for the on-shell action $\cs$, namely
\bea\label{constraints}
&&{\rm Vol}^{-2}\,\Big[\frac12\h^{ij}\frac{\d \cs}{\d X^i}\frac{\d \cs}{\d X^j}
-\frac{1}{2\vecX^2}\Big(\frac{\d\cs}{\d \f}\Big)^2\Big]-2+V(\f)=0\\
&&{\rm Vol}^{-1}\,\e_{i}{}^{jk}X^i\frac{\d \cs}{\d X^j}=J^k.
\eea 
Hamilton-Jacobi theory is based on the fact that finding a complete integral (i.e.~a solution that contains 
as many integration constants as there are fields) of these PDEs is equivalent to 
completely solving the EOM. Mathematically, this follows from the theory of first order PDEs 
and the fact that Hamilton's equations are the characteristic equations of the Hamilton-Jacobi PDEs.    

The boundary term that is required to render the on-shell action finite agrees with a suitable solution
of the Hamilton--Jacobi equations up to terms that vanish at infinity \cite{de Boer:1999xf}. In fact, this 
same boundary term is required to make the 
variational problem well posed \cite{Papadimitriou:2005ii,Papadimitriou:2010as}. As we shall see, 
not any solution of the Hamilton--Jacobi equations can be used in the boundary counterterms, but the suitable
solution is not unique either in general. The non-uniqueness originates in the freedom of choosing the integration
constants in the solution of the Hamilton--Jacobi equations. These integration constants only affect the finite part
of Hamilton's principal function. The holographic interpretation of this ambiguity is the usual renormalization
scheme dependence
of the renormalized generating functional in the dual field theory. This observation is particularly important for
the problem at hand, because the angular momentum $J_i$ corresponds to one of these integration constants and so it
only affects the finite part of the on-shell action. For the purposes of determining the boundary counterterms,
therefore, it suffices to consider only zero angular momentum solutions of the Hamilton--Jacobi equations. Any such
solution takes the form  
\beq\label{zero-j-solution}
\cs_{J=0}={\rm Vol}\;2W(\f)\sqrt{\vecX^2},
\eeq
where $W(\f)$ is a function that is related to the potential by
\beq\label{potential-eq}
V(\f)=2+2W'^2-2W^2.
\eeq
The boundary term in \eqref{eq:quench17} is exactly of the form \eqref{zero-j-solution}, but with $W(\f)$ 
replaced with the function $U(\f)$. Note that $W(\f)$ and $U(\f)$ are {\em not} the same function, and this is
why we use different symbols. $U(\f)$ is the function that defines the boundary counterterms, while $W(\f)$ 
denotes any exact solution of \eqref{potential-eq}, i.e.~the so called `fake superpotential'. $U(\f)$ is in fact
related to a {\em particular} solution $W(\f)$ in a way we will describe below.   

Before we analyze the various solutions of \eqref{potential-eq} for the scalar potentials we presented in the
previous section, it is important to realize that the solutions $W(\f)$ of this equation determine 
the asymptotic form of the fields themselves via the first order equations 
\beq\label{flow-eqs}
\dot X^i=\frac{2W(\f)}{\sqrt{\vecX^2}}X^i \qquad \dot\f=-\frac{2W'(\f)}{\sqrt{\vecX^2}}\,
\eeq
which are obtained by combining the expressions \eqref{momenta} and \eqref{HJ-momenta} for the canonical momenta.
Since the point particle angular momentum $J_i$ corresponds to a normalizable mode of the metric, the asymptotic
expansions obtained through \eqref{flow-eqs} will be correct up to the order of this mode.  
These flow equations, therefore, apart from helping in determining which solution of \eqref{potential-eq} should 
be used in the counterterms, also relate the coefficients in the Taylor expansion of the potential to the 
coefficients of the asymptotic expansion of the input function $f_1(\r)$.

For $0<\ve<1$, the solutions of \eqref{potential-eq} with the potential \eqref{general-ads-potential} take one 
of the two possible asymptotic forms
\bea\label{superpotential-1}
W(\f)=\left\{\begin{matrix}
1+\frac{\D_+}{4}\f^2+\cdots\\
1+\frac{\D_-}{4}\f^2+w_3\f^3+w_4\f^4+\cdots+w_n\f^{n}+\tilde{w}_{n+1}\f^{n+1}\log\f+\l \f^{\frac{2}{\D_-}}
+\cdots
\end{matrix}\right.
\eea
where the coefficients $w_k$ are related to the coefficients $v_k$ in the potential, except for the constant 
$\l$ which is an integration constant of the first order equation \eqref{potential-eq}. 
The dots in \eqref{superpotential-1} 
stand for subleading terms that do not contribute to the on-shell action once the radial regulator is removed.
Moreover, the logarithmic term can be present only in the case of resonant scalars \eqref{eq:resonant}. 
The function $U(\f)$ that defines the counterterms must, therefore, take one of these two forms, with the terms 
denoted by dots dropped since they do not contribute to the on-shell action.  
To determine which of the two asymptotic forms $U(\f)$ must take, notice that the coefficient of $\f^2$ in $W(\f)$
determines, via the first order equations \eqref{flow-eqs}, the leading asymptotic behavior of the corresponding solutions for $\f$. Namely, the two possible signs lead respectively to  
\beq
\f\sim \r^{-\D_\pm/2}. 
\eeq
These are precisely the asymptotic forms of the two possible independent solutions for 
the scalar field in \eqref{eq:quench71} and (\ref{scalargeneral}). The plus sign, therefore, corresponds to 
solutions of the EOMs with $\f_-=0$, i.e.~Brown-Henneaux boundary conditions. Hence, in order to ensure that 
the boundary term makes the on-shell action finite on all possible solutions the function $U(\f)$ must be of
the second form in \eqref{superpotential-1} \cite{Papadimitriou:2004rz}. Moreover, locality of the boundary term
means that the term proportional to $\l$ can be included only when the exponent $2/\D_-$ is integer, i.e.~only
for resonant scalars \eqref{eq:resonant}. In that case, $\l$ can be non-zero and its value can be chosen at will as this term only
changes the finite value of the renormalized on-shell action and of the other renormalized quantities. 
It corresponds to part of the renormalization scheme dependence of the dual theory. There is a unique value, 
$\l_*$, of $\l$ in $U(\f)$ for which the renormalized stress tensor in the case of resonant scalars remains of the 
form \eqref{stresstensor}, and we assume that this choice of scheme is made.  
Finally, the logarithmic term in the case of resonant scalars would seem to make the boundary term non-local in $\f$.
This is a typical signature of a conformal anomaly. Locality can be preserved at the expense of introducing
cut-off dependence in the boundary counterterms. Since $\f\sim \r^{-\D_-/2}$, putting everything together, 
we conclude that for $0<\ve<1$ the function $U(\f)$ in \eqref{eq:quench17} takes the from 
\beq\label{U-function-1}
U(\f)=1+\frac{\D_-}{4}\f^2+w_3\f^3+w_4\f^4+\cdots+w_n\f^{n}-\frac{1-\ve}{2}\tilde{w}_{n+1}\f^{n+1}\log\r_\infty
+\lambda_* 
 \f^{\frac{2}{\D_-}}
\eeq
where $\r_\infty$ is the radial cut-off and the last two terms are present only in the case of resonant scalars. 
The coefficients $w_k$ and $\tilde w_{n+1}$ can be related via the first order equations \eqref{flow-eqs} to the
coefficients of the asymptotic expansion for the input function $f_1(\r)$. For the first couple of terms we find
\beq
w_3=\frac{1}{16}(3\ve-1)a_3 \quad w_4=\frac16(2\ve-1)a_4 -\frac{15(3\ve-1)^2}{256(1-\ve)}a_3^2+\frac{(1-\ve)^2}{48}
\eeq 
while
\beq
\tilde w_{n+1}=-\ve\tilde\m_{n+1}.
\eeq
Note that the counterterms \eqref{U-function-1} are local in the induced fields,
which ensures that the dual CFT is a local renormalizable field theory. It was precisely this requirement 
that led us to the form \eqref{general-ads-potential} of the scalar potential.

For $\ve=0$, the potential \eqref{BF-potential} leads to the two asymptotic solutions 
of \eqref{potential-eq} 
\cite{Papadimitriou:2007sj}
\beq\label{superpotential-2}
W(\f)=\left\{\begin{matrix}
1+\frac14\f^2+o\left(\frac{\f^2}{\log\f}\right) \\
1+\frac14\f^2\left(1+\frac{1}{\log\f}\right)+o\left(\frac{\f^2}{\log\f}\right)
\end{matrix}\right.
\eeq
Again, the first solution corresponds to the mode $\f_-$ in \eqref{BF-expansion} being switched off and, hence,
the function $U(\f)$ must correspond to a solution of the second type in \eqref{superpotential-2}. 
As in the case of the resonant scalars, in order to preserve locality we must introduce explicit cut-off 
dependence in the boundary counterterms. Since $\log\f\sim -\frac12\log\r$, we deduce that 
the counterterm function for $\ve=0$ takes the form \cite{Bianchi:2001kw,Papadimitriou:2004rz}    
\beq
U(\f)=1+\frac14\f^2\Big(1-\frac{2}{\log\r_\infty}\Big)\,,
\eeq
where $\r_\infty$ is the location of the cut-off.

Finally, for $\ve=1$ ($V(\phi)=0$) the scalar does not contribute to the divergences of the on-shell action 
and one only needs to remove the volume divergence by choosing in \eqref{potential-eq} the solution 
\beq
U(\f)=1\,.
\eeq


\section{Black hole thermodynamics}\label{se:TD}

Black hole thermodynamics provides some key insights into semi-classical and quantum gravity. Of particular importance is the black hole entropy and its microscopic description, for instance from a CFT perspective by virtue of the Cardy formula. In this section we address these issues for EDG.

In section \ref{se:6.1} we discuss basic thermodynamical quantities, like temperature, entropy, etc. We derive a formula for entropy in terms of curvature invariants.
In section \ref{se:6.2} we focus on black hole families and their associated solitons. We prove that these solitons can never have a (free) energy lower than global AdS for marginal boundary conditions.
In section \ref{se:6.3} we show the validity of the Cardy formula in EDG, again for marginal boundary conditions.

\subsection{Basic thermodynamical quantities}\label{se:6.1}

Black hole thermodynamics can be studied efficiently on the gravity side in the Euclidean path integral approach, by exploiting results from the previous sections. We refer to \cite{Papadimitriou:2005ii} for a general analysis of black hole thermodynamics in the context of holography in asymptotically locally AdS backgrounds. This section is a particular application of the results there to the 3-dimensional case we are discussing here, taking into account the modifications required by the possibility of generalized boundary conditions. See also \cite{Grumiller:2007ju} 
for a similar analysis of black hole thermodynamics in 2-dimensional EDG.  

\paragraph{Temperature}
Demanding the absence of a conical defect at the horizon $\rho\!=\!\rho_h$ leads to a periodicity in Euclidean time $\tau\sim\tau+\beta$ that is identified with the inverse temperature. From the line-element \eqref{eq:ADM} we obtain
\bea
\b=T^{-1}={4 \pi \over R \, \partial_\rho N^2}\, \Big|_{\rho_h}= {2 \pi  R_h \over j}  \,.
\label{eq:quench199}
\eea
Notice that for black hole solutions the particle angular momentum $j$ is positive.

\paragraph{Entropy}
To derive entropy from scratch we could evaluate first the on-shell action, extract from there the free energy and then obtain entropy by taking the appropriate partial derivative of the free energy with respect to temperature. We shall provide these results below.
Alternatively, for a two-derivative gravity theory it is known that the black hole entropy simply is given by the Bekenstein-Hawking formula 
\beq
S=4\p A_h 
\label{eq:BH}
\eeq
where $A_h\!=\!2 \pi R_h$ is the horizon area. 
Using \eqref{eq:quench199}, an interesting reformulation of the result for entropy is
\eq{
S = 2\,\textrm{Vol}_E \, j
}{eq:quench178}
where $\textrm{Vol}_E\!=\!\int \! \extd\tau \extd \vf$. The entropy formula \eqref{eq:quench178} highlights the physical importance of the constant of motion $j$ from \eqref{eq:quench12}.

Our result \eqref{eq:quench178} allows a relation between entropy and the ratio of curvature invariants, the square of the Cotton tensor and the square of the tracefree Ricci tensor, that resembles Penrose's Weyl curvature conjecture in four dimensions \cite{Penrose:1977}:
\eq{
S = {\cal A}\,\sqrt{\frac{C_{\mu\nu} C^{\mu\nu}}{\slash R_{\mu\nu}\slash R^{\mu\nu}}}
}{eq:quench179} 
The prefactor is given by ${\cal A}=\tfrac{4}{\sqrt{3}} 
\,\int\extd^2x\sqrt{-\ga}$, where the integral goes over some $\rho=\rm const.$~hypersurface with induced metric $\gamma$. The relation \eqref{eq:quench179} is independent from the particular choice of hypersurface and can be checked easily using \eqref{eq:quench21}, \eqref{eq:quench26} and \eqref{eq:quench178}.

\paragraph{Angular velocity}
Since a black hole is a localized object, its angular velocity is measured with respect to that at infinity, which is a property of the asymptotic region
\beq
\omega=N_{\vf \, h}-N_{\vf \, \infty}\,.
\label{eq:quench198}
\eeq
In the AdS case we are currently considering and without loss of generality, we then fix the asymptotic form of the spacetime by requiring the boundary metric to be Minkowski. Namely, $N_{\vf \, \infty}\!=\!0$ and $R^2_{\infty}\!=\!2$. This reduces the map between the local and physical coordinates \eqref{eq:quench137} to \eqref{eq:eta}, as was already used for the derivation of BTZ solutions. This restriction yields
\eq{ 
\omega = \tanh \eta
}{eq:add100}
with $\eta$ as defined in \eqref{eq:eta}.

\paragraph{Mass and angular momentum} 
The holographic conserved charges associated with a boundary conformal Killing vector $k^i$ are given by
\eq{
\cq[k]=\int_0^{2\pi}\limits  \extd\vf \,\langle T_{ti}\rangle\, k^i\,.
}{eq:add101}
Since we know the renormalized holographic stress tensor, it is straightforward to evaluate these
conserved charges. 
The black hole mass is obtained for $k\!=\!\partial_t$
\beq
M=\int\limits_0^{2\p}\extd\vf\, \big( \langle T_{00} \rangle \cosh^2 \eta + \langle T_{11} \rangle \sinh^2 \eta)
\eeq
where the stress tensor should be evaluated with the appropriate boundary conditions, and the subindices here refer to the local coordinates $(x^0,x^1)$. Substituting \eqref{stresstensor} for Dirichlet, or  \eqref{eq:stress-tensor-vev} for Neumann or Mixed boundary conditions, we obtain
\eq{
M=2\pi j \cosh 2\eta - 2\pi \langle T^i_i \rangle\,.
}{eq:add102}  
The black hole angular momentum is obtained for $k\!=\!\partial_\vf$.
\beq
J=\int\limits_0^{2\p}\extd\vf\,\langle T_{t\vf}\rangle = \pi \,\big(\langle T_{00}\rangle+\langle T_{11}\rangle\big) \sinh{2\eta}= 2\pi j \sinh{2\eta}
\eeq

\paragraph{Gibbs free energy and the first law of black hole mechanics} 
The renormalized Euclidean on-shell action gives the Gibbs free energy \cite{Gibbons:1977ue} 
\beq
I=\b G(T,\omega)
\eeq
where $I=-\Ga$ is the renormalized Euclidean on-shell action and 
\beq
G(T,\omega)\equiv M -TS-\omega J
\label{eq:add106}
\eeq
is the Gibbs free energy. This coincides with the Helmholtz free energy for vanishing angular velocity and 
angular momentum. 
Evaluating the Euclidean on-shell action we get
\beq
 G(T,\omega)=-2\pi j  -2 \pi \langle T^i_i \rangle\,.
\eeq
Indeed, using the relations derived above, this expression reproduces the right hand side of \eqref{eq:add106}, and $-\partial G/\partial T|_{\om}$ reproduces the Bekenstein--Hawking entropy \eqref{eq:BH}.

Finally, as is shown in general in \cite{Papadimitriou:2005ii} these quantities satisfy the first law of
black hole mechanics
\beq
\label{eq:first-law}
\d M=T\d S+\omega \d J
\eeq
where the variations are taken at fixed source $\cj$ and fixed boundary condition for the scalar field. 

\subsection{Black hole families and their solitons}\label{se:6.2}

We have seen in previous sections how to systematically construct black hole families by varying, besides the rotation parameter $\eta$, the integration constant $\Om$. We analyze here their thermodynamic properties. 

The black hole temperature, \eqref{eq:quench199}, is
\eq{
T= {1 \over 2 \pi \cosh \eta} \sqrt{\Om \over \, \Om_s}
}{eq:add111}
with $\Om_s$ the value of $\Om$ for the unique static soliton solution associated to each black hole family, given in \eqref{eq:add009}.  As we have discussed, only marginal boundary conditions with vanishing source allow to freely vary the parameter $\Om$. Namely, this is the boundary condition for which the $\{\Om,\eta\}$-black hole family can be interpreted as thermal states in the same boundary CFT. Since for marginal boundary conditions the trace of the stress tensor is proportional to the sources, \eqref{eq:add104}, the field theory energy and angular momentum are given by
\eq{
M=8\pi  \Om   \cosh 2\eta \qquad J=8\pi  \Om   \sinh 2\eta
}{eq:add112}  
We have chosen $\Om$ satisfying \eqref{eq:add200}, such that $j\!=\!4\Om$ for simplicity. The entropy is
\eq{
S=32 \pi^2 \cosh \eta\, \sqrt{\Om_s \Om}\,.
}{eq:add114}
The charges of the solitonic solution are
\eq{
M_s=-8\pi\,  \Om_s   \qquad J=0
}{eq:add120}  


Any of these 2-parameter families of stationary hairy black holes includes a 1-parameter family of extremal solutions. It is obtained by performing the double scaling limit $\Om\!\rightarrow \!0$, $|\eta|\!\rightarrow\!\infty$ while keeping fixed $\Om \,e^{2 |\eta|}\!\rightarrow \!{\bar \Om}$. We then have
\eq{
M=|J|=4\pi  {\bar \Om}   \qquad S=16 \pi^2 \sqrt{\Om_s {\bar \Om}} \qquad T=0\,.
}{eq:add115}  

\paragraph{Stability of the hairy sector}

In addition to solutions with non-trivial scalar profile, the boundary condition \eqref{eq:add201} clearly allows solutions with vanishing scalar: the BTZ black hole family and its associated soliton, global AdS. The preferred solution will be that with smaller free energy for a given temperature. 

The free energy for hairy or BTZ black holes and their respective solitons is given by
\eq{
G = -8\pi\, \Om \, \qquad G_s=M_s=-8 \pi\, \Om_s\,.
}{eq:add500}
A first interesting implication is that inside each family there is a Hawking-Page phase transition at $T\!=\!\tfrac{1}{2\pi}$. For $T\!<\!\tfrac{1}{2\pi}$, or equivalently $\Om\!<\!\Om_s$, the soliton geometry with a thermal circle is preferred, while for $T\!>\!\tfrac{1}{2\pi}$ the preferred one is the black hole. Comparing members of the hairy and BTZ families with the same temperature \eqref{eq:add111}, we obtain
\eq{
G={\Om_s \over \Om_{\textrm{\tiny AdS}}}\, G_{\textrm{\tiny BTZ}}   \qquad M_s={\Om_s \over \Om_{\textrm{\tiny AdS}}}\, M_{\textrm{\tiny AdS}}
}{eq:add501}
Which family dominates just depends on the quotient $\Om_s/\Om_{\textrm{\tiny AdS}}$. 

We show now that $\Om_s\!<\!\Om_{\textrm{\tiny AdS}}$ for any smooth solution with a center in EDG with non-trivial scalar profile. Relation \eqref{eq:add0} can be rewritten as
\eq{
\Om_s={1 \over 2 {\dot f}_{1c}}
}{eq:add502}
where the subindex ``c'' indicates that the function should be evaluated at the center. We are assuming $f_1$ to be asymptotically positive, in particular $\dot f_1\!=\!2$ at infinity. The inequality \eqref{eq:quench14} requires then its second derivative to be negative for all values of the radial coordinate. Hence $\dot f_1$ is a decreasing function, and we obtain
\eq{
\Om_s\leq{1 \over 4}\,.
}{eq:add503}
The inequality is saturated only when $\ddot f_1$ vanishes for all radial values, which implies that the solution is global AdS. We can state the following interesting result: any smooth solution of EDG with vanishing trace of the boundary stress tensor has bigger mass than global AdS.

Given that the free energy \eqref{eq:add500} is negative, the previous result ensures that a BTZ black hole has smaller free energy than the corresponding hairy one (for marginal boundary conditions). In spite of that, these hairy black holes are thermodynamically perturbatively stable since their specific heat at constant angular potential is always positive:
\eq{
C = T\,\frac{\partial S}{\partial T}\Big|_{\omega} = S \geq 0
}{eq:quench144}
This further implies that logarithmic corrections to the entropy from thermal fluctuations in the large $\Omega$ limit take the form $S_{\rm tot}=S+\tfrac32\,\ln{S}+\dots$, which is the same result as for the BTZ black hole, see e.g.~\cite{Sen:2012dw} and references therein.

We would like to stress, however, that thermodynamic stability of a solution depends strongly on the boundary conditions. The same solitonic solutions may appear in other theories with different boundary conditions. The soliton mass $M_s$ depends on them through a contribution from the trace of the boundary stress tensor, see \eqref{eq:add102}. If that boundary conditions admit also global AdS as a solution, the soliton mass might be lower than that of AdS. Due to the inequality
\eq{
M_s = -8 \pi \Om_s  -2\pi\, \langle T^i_i\rangle \geq M_{\textrm{\tiny AdS}} - 2\pi\,\langle T^i_i\rangle
}{eq:nogo3}
these solitonic solutions necessarily must have a positive trace of the boundary stress tensor, $\langle T^i_i\rangle > 0$. Finally, even for marginal boundary conditions note that the results above do not automatically imply that soliton solutions will tunnel into the AdS vacuum, since these vacua might be disconnected. 

\subsection{Cardy formula}\label{se:6.3}

The Cardy formula for specific hairy black holes in EDG was considered in \cite{Park:2004yk,Correa:2011dt}.
\eq{
S_{\textrm{\tiny Cardy}}=4\pi \sqrt{-\De_0^+ \De^+} + 4\pi\sqrt{-\De_0^-\De^-}
}{eq:cardy}
where $\De^\pm=(M\pm J)/2$ are the Virasoro zero mode eigenvalues of the black hole states, and $\De_0^\pm$ are the lowest eigenvalues. The non-trivial question is what those lowest eigenvalues should be. 
Rewriting the entropy formula \eqref{eq:add114} in a suggestive way,
\eq{
S=2\pi \sqrt{-M_s (M+J)} + 2\pi \sqrt{-M_s (M-J)}\,,
}{eq:add116}
we see that the Cardy formula is satisfied, $S=S_{\textrm{\tiny Cardy}}$, provided we identify $\De_0^\pm \!=\!\tfrac12\,M_s$, as proposed in \cite{Correa:2011dt}. Since our conclusions hold for any $(\Om,\eta)$-family of stationary black hole solutions we have generalized the results by Correa, Martinez and Troncoso from a set of specific examples to generic EDG.

The fact that a Cardy formula holds for hairy black holes is remarkable and may indicate that the hairy sector is actually stable against tunneling into the BTZ sector.


\section{Examples}\label{se:5}
In this section we present some  examples to illustrate our algorithm.
In section \ref{se:5.1} we outline (and give explicit examples) how to obtain specific solutions that appeared in the literature on 3-dimensional gravity with a self-interacting scalar field \cite{Henneaux:2002wm,Clement:2003sr,Gegenberg:2003jr,Henneaux:2004zi,Correa:2010hf,Correa:2011dt,Correa:2012rc}.
In section \ref{se:5.2} we present solitonic solutions that obey Brown--Henneaux boundary conditions.
In section \ref{se:5.3} we discuss a class of asymptotically AdS examples that violate Brown--Henneaux boundary conditions, for any scalar field mass in the window $-1<m^2<0$.
In section \ref{se:2.5} we consider solutions with exponential potential, which generally leads to solutions that are neither asymptotically flat nor asymptotically (A)dS.

\subsection{Recovering known solutions}\label{se:5.1}

By use of our algorithm we should be able to construct all stationary axi-symmetric solutions. In particular it should reproduce all solutions in the literature, most of which are static ones. To check this, given a static solution
\beq
\extd s^2 = g_{tt} \extd t^2 + g_{\varphi\varphi} \extd\varphi^2 + g_{rr} \extd r^2 \label{genericmetric}
\eeq
there exists a coordinate $\rho$ such that the metric looks like our \eqref{eq:quench19}.
Taking $f_2 = -g_{tt}$, $f_1 = g_{\varphi\varphi}$ it follows that the required coordinate change is
\beq \label{g}
g(r) \equiv \frac{\extd r}{\extd\rho} = \frac{1}{\sqrt{-g_{tt}g_{\varphi\varphi}g_{rr}}}\,.
\eeq
The EOM were written in the $\rho$ coordinates. The explicit expression for $r(\rho)$ might be hard to obtain, therefore  we convert the relevant $\rho$-coordinate EOM into $r$ coordinates (primes denotes derivatives with respect to $r$).
\bea
\phi' = -\sqrt{-\frac{f_1''+ \frac{g'}{g} f_1'}{f_1}}=-\sqrt{-\frac{f_2''+ \frac{g'}{g} f_2'}{f_2}} 
\eea
Under the same change of coordinates, the integration constant $j$ is
\beq
2j = g(r)\,\big(f_2'(r)f_1(r)-f_1'(r)f_2(r)\big)\,.
\eeq

As examples we discuss in detail how to recover the solutions from \cite{Henneaux:2002wm} and \cite{Gegenberg:2003jr}.
The metric is given by
\beqa
g_{tt} &=& - \frac{H^2}{(H+B)^2}  F(r)\,, \qquad g_{rr} = \frac{(H+B)^2}{(H+2B)^2 F(r)}  \,,\qquad g_{\varphi\varphi} = r^2\,,
\eeqa
where
\beqa
H(r) &=& \frac{1}{2}(r+\sqrt{r^2+4Br}),\qquad
F = \frac{H^2}{l^2} -(1+\nu)\Big( \frac{3B^2}{l^2}+\frac{2B^3}{l^2H} \Big)\, .
\eeqa
Using the coordinate transformation (\ref{g}),
\beq\label{trafo1}
g(r)=\sqrt\frac{4B+r}{r^3}\,,
\eeq
the scalar field is then determined as
\beq
\phi = 4\,\arctanh\sqrt{\frac{B}{H(r)+B}}\,.
\eeq
Taking into account the different normalizations,  $4\Phi=\phi$, where $\Phi$ denotes the scalar field used in \cite{Henneaux:2002wm}, we arrive at the potential
\beqa\label{potential1}
V(\Phi) = -\frac{1}{8l^2}\, \big(\cosh^6 \Phi + \nu \sinh^6 \Phi\big) \label{potential15} 
\eeqa
Furthermore, we find the relation between constants of motion and parameters
\beq
j = 3 B^2(1+\nu). \label{j}
\eeq
Since $\nu$ is part of the potential and $B^2$ is not, the quantity $B^2$ plays the role of our constant of motion $j$, whereas $\nu$ is a free parameter of the model.

Now we check that our general results for physical observables like mass and entropy reproduce the results obtained in \cite{Henneaux:2002wm} through the formalism of asymptotic symmetries. 
To this end we write the action of \cite{Henneaux:2002wm} in a way in which the different normalizations are more apparent:
\beqa
S &=& \frac{1}{\pi G} \int d^3x \sqrt{-g} \left[ \frac{R}{16} -\frac{1}{2}(\nabla \Phi)^2 - V(\Phi) \right] \\
&=& \frac{1}{16\pi G}\int d^3x \sqrt{-g} \left[ R +2 - \frac{1}{2}(\nabla \phi)^2 -V(\phi) \right]
\eeqa
where we defined
\beq
V(\phi) =  -\frac{3}{8}\phi^2 -\frac{1}{32}\phi^4 -\frac{47+15\nu}{30720} \phi^6 -\dots
\eeq
which corresponds to $\varepsilon=\tfrac12$ (resonant scalars) according to (\ref{eq:quench72}) and (\ref{eq:angelinajolie}).
Note that the two leading terms show a universal behavior expected from \eqref{eq:nolabel2} for $\eps=\tfrac12$ and $a_3=0$.
As explained in detail in section \ref{se:BF} observables of interest depend on the chosen boundary conditions (\ref{general-bc}).
In order to calculate the mass we need to write $\phi(r)$ as in \eqref{eq:quench71}
and then identify a function $\mathcal W'(\phi_-)$ such that the regularity conditions (\ref{mixed-ct}) holds.
In the case at hand  $\varepsilon = 1/2$ and therefore  $\Delta_- = 1/2$ and $\Delta_+ = 3/2$.

Since we are only interested in the asymptotics it is sufficient to take the leading terms of $\rho=\rho(r)$.
Expanding and integrating (\ref{trafo1}) leads to 
\beq
\frac{r^2}{2} - 2B r + \cdots = \rho\,.
\eeq
Inverting the above expression and substituting into  \eqref{eq:quench71} gives
\beq
\phi(\rho) =  2^{7/4}  \sqrt{B} \rho^{-1/4} -\frac{10}{3} 2^{1/4} B^{3/2} \rho^{-3/4} + \dots
\eeq
so that we can identify $\phi_- = 2^{7/4} \sqrt{B}$ and $\phi_+ = -\frac{10}{3} 2^{1/4} B^{3/2}$. Using (\ref{mixed-ct}) we further identify
\beq
\mathcal W'(\phi_-) = -\frac{5}{48} \phi_-^3 \,,
\eeq
or $\mathcal W(\phi_-) = -\frac{5}{192} \phi_-^4 + w$. Equivalently $\mathcal W(\phi_-) = -\frac{10}{3} B^2 + w$. The integration constant $w$ corresponds to an unphysical additive constant of the effective potential \eqref{effective-potential}.
Assuming Dirichlet boundary conditions the renormalized stress energy tensor (\ref{stresstensor}) is
\beq
\langle T_{00} \rangle_+ = \frac{B^2}{3} \big( 9(1+\nu) +20 \big)\,.
\eeq
For mixed boundary conditions it follows from   (\ref{eq:stress-tensor-vev}) that 
\beq
\langle T_{00} \rangle_- = 3 B^2(1+\nu) +2 w\,.
\eeq
To obtain the mass we integrate in $\varphi$ (multiply by $2\pi$) and use the reference's different normalization (divide by $16\pi$), and obtain
\beq
M = \frac{3}{8} B^2(1+\nu) +\frac{1}{4} w
\eeq
which agrees with the result of  \cite{Henneaux:2002wm} for $w=0$. 

There exits a second class of solutions obtained from the potential (\ref{potential1}) for $\nu\le0$ presented in \cite{Gegenberg:2003jr}. The metric is given by
\beqa
g_{tt} &=& -r^2, \\
g_{\varphi\varphi} &=& r^2 ,\\
g_{rr} &=& \frac{(f+B(\sqrt{-\nu}-1))^2(f+B\sqrt{-\nu})^2}{f^2(f^2-B(B-2f)\sqrt{-\nu}-B^2\nu)^2} \\
f(r) &=& \frac{1}{2}\Big( r-B(\sqrt{-\nu}-1)+\sqrt{(r-B(\sqrt{-\nu}-1))^2+4B\sqrt{-\nu}r}\Big)\,.
\eeqa
We again transform the coordinates via (\ref{g}). The scalar field asymptotically is given by
\beq
\phi(r) = 4\sqrt{\frac{B}{r}} + \big(-\tfrac{2}{3}-2\sqrt{-\nu} \big) \Big(\frac{B}{r}\Big)^{3/2} + \dots
\eeq
and agrees with
\beq
\Phi(r) = \arctanh \sqrt{\frac{B}{f+B\sqrt{-\nu}} }\,, \\
\eeq
by taking into  account the different conventions, $4\Phi=\phi$. Note that in this case $j=0$ and 
we have $m^2 = -3/4$, $\Delta_+=3/2$ and $\Delta_-=1/2$ as before.
The change of variables from $r$ to $\rho$ is (perturbatively)
\bea
g(r) &=& \frac{1}{r} + \frac{2B}{r^2} + \cdots\;, \\
r &=& 2B + \sqrt{2}\sqrt{2B^2+\rho}\;, \\
\phi(\rho) &=& 2^{7/4} \sqrt{B} \rho^{-1/4} -\frac{1}{3}(2^{1/4}B^{3/2}(7+3\sqrt{-\nu}))\rho^{-3/4}\;,
\eea
leading to
\bea
\phi_+ &=& -\frac{1}{3}(2^{1/4}B^{3/2}(7+3\sqrt{-\nu})) \;,\\
\phi_- &=& 2^{7/4} \sqrt{B}\;.
\eea
Through the regularity condition (\ref{mixed-ct}) we obtain
\bea
\mathcal W'(\phi_-) &=& -\frac{1}{96}(7+3\sqrt{-\nu}) \phi_-^3 \\
\mathcal W(\phi_-) &=& -\frac{1}{384}(7+3\sqrt{-\nu}) \phi_-^4 + w\,.
\eea
Inserting everything into \eqref{eq:stress-tensor-vev} gives
\bea
\langle T_{00} \rangle_+ &=& \frac{2}{3} B^2 (7+3\sqrt{-\nu}) \\
\langle T_{00} \rangle_- &=& \langle T_{00} \rangle_+ + 2 \mathcal W(\phi_-) = 2w
\eea
in agreement with \cite{Gegenberg:2003jr}. Because the stress tensor does not depend  on $j$ (or equivalently $B$), this is called a degenerate state.

\subsection{Brown--Henneaux solitons}\label{se:5.2}

Here we focus on asymptotically AdS solutions with specifically designed properties.
We are particularly interested in regular solutions without black hole horizons.
There are two possibilities to achieve this: 
either we make sure there is no horizon but rather a center, or we ensure that the horizon corresponds to a Poincar\'e patch horizon and not a black hole horizon.
We start with the latter option.

\paragraph{Poincar{\'e} patch soliton}
As discussed in section \ref{se:2.2} static solutions exhibit a Poincar\'e patch horizon if the functions $f_1$ and $f_2$ simultaneously have a zero.
This can be achieved most easily if both functions coincide with each other.
A simple representative choice is
\eq{
f_1=f_2=2\rho\,\Big(1+\frac{\Om}{\rho+\Om}\Big)\,,
}{eq:quench90a}
with some positive $\Om$.
This choice ensures asymptotic AdS behavior with Brown--Henneaux boundary conditions for large $\rho$ and a Poincar\'e patch horizon at $\rho=0$.
The scalar field \eqref{eq:quench13} is given by 
\eq{
\sqrt{2}\phi = \pi-4\arctan\sqrt{\frac{\rho}{\rho+2\Om}}
}{eq:quench92}
where we chose the integration constant such that $\lim_{\rho\to\infty}\phi=0$.
At the Poincar\'e patch horizon the scalar field approaches the value $\lim_{\rho\to 0}\phi=-\pi/\sqrt{2}$.
Evaluating the scalar potential \eqref{eq:quench15} yields a simple result.
\eq{
V(\sqrt{2}\phi)= - 6\sin^4\!\phi
}{eq:quench94}

\begin{figure}
\begin{center}
\epsfig{file=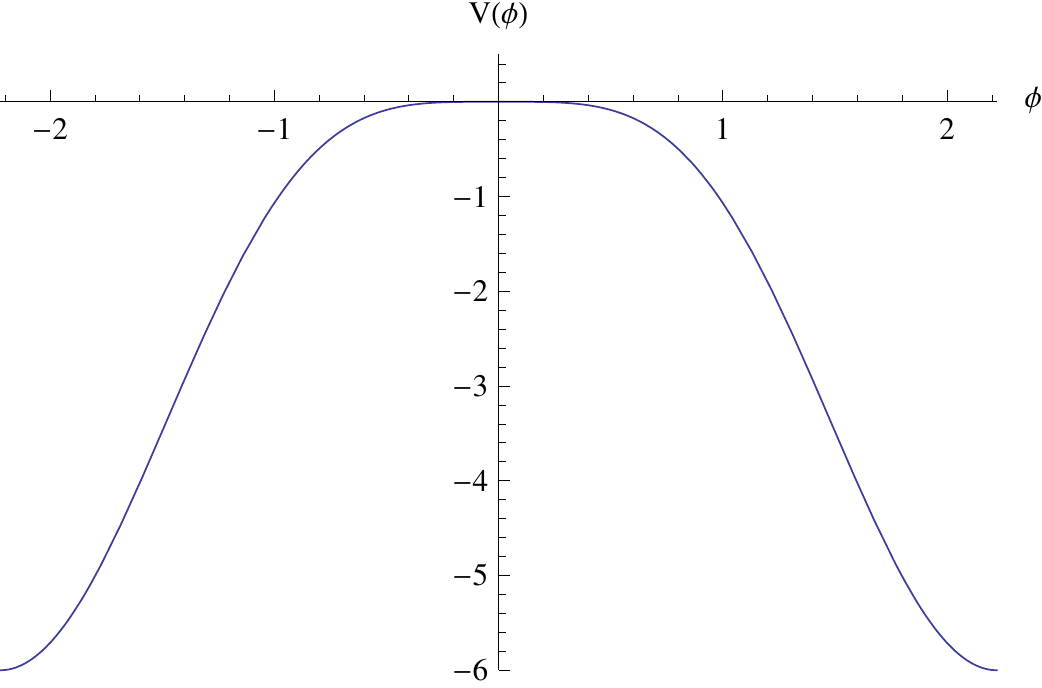,width=0.4\linewidth}
\end{center}
\caption{Potential $V(\phi)$ for (7.37)}
 \label{fig:2}
\end{figure}

Figure \ref{fig:2} depicts the potential \eqref{eq:quench94} in the whole range of $\phi$.
The asymptotic AdS region corresponds to the maximum of the potential in the center, while the minimum corresponds to the Poincar\'e patch horizon.
The solutions following from the Ansatz \eqref{eq:quench90a} are a special case ($j=0$) of the next family of solutions.

\paragraph{Soliton with center} As discussed in section \ref{se:2.2} static solutions exhibit a regular center if the function $f_1$ has a zero, but $f_2$ has no zero (or vice versa if the meaning of time and angular coordinate are exchanged).
Let us start by choosing again
\eq{
f_1=2\rho\,\Big(1+\frac{\Om}{\rho+\Om}\Big)\,.
}{eq:quench90}
Then we can integrate the ODE \eqref{eq:quench12} and obtain
\eq{
f_2 = f_1 - \frac{\Om\,\xi}{2}\, \Big[1+\frac{\rho(\rho+2\Om)}{2\Om(\rho+\Om)}\,\ln\Big(1+\frac{2\Om}{\rho}\Big)\Big]\,.
}{eq:quench91}
For $\xi=0$ the Poincar\'e patch soliton of the previous paragraph is obtained.
If $\xi$ is positive the function $f_2$ has a zero at some positive value of $\rho$.
If $\xi$ is negative the function $f_2$ has no zero in the whole domain $\rho\in[0,\infty)$ and regular solitons with center are obtained.

The scalar field is given by \eqref{eq:quench92}.
Evaluating the scalar potential \eqref{eq:quench15} generalizes the result \eqref{eq:quench94} to something more complicated.
Consistently with our general discussion, the potential \eqref{eq:quench94} is independent from the constant of motion $\Om$, but like in the asymptotically flat example in section \ref{se:3} it does depend on $\xi$.

It is straightforward to generalize the discussion above to functions with similar behavior, for instance $f_1=2\rho\,\big(1+\Om/(\rho+\Om a^2)\big)$, with some parameter $a$. The general features remain the same, but the potential is more complicated, which is why we refrain from presenting further explicit examples of this type.

\subsection{Non-Brown--Henneaux solutions}\label{se:5.3}

An interesting class of input functions $f_1$ that asymptotically behave as (\ref{eq:mostgen}) for $0\leqslant\epsilon<1$ is given by 
\begin{equation}\label{Besself1}
f_1(\rho)=N_1\,\sqrt{\Om\rho}\,\mathrm{Y}_{\frac{1}{1-\eps}}\left(\frac{\rho}{\Om}\right)^{\frac{\eps-1}{2}}
\end{equation}
with $N_1=-2^\frac{\eps}{\eps-1}\pi/\Gamma\left(\frac{1}{1-\eps}\right)$.
The advantage of the choice above is that we can exploit properties of Bessel functions Y$_n$ and J$_n$.
Integrating the ODE (\ref{eq:quench12})  leads to
\begin{equation}\label{Besself2}
f_2(\rho)=f_1(r)+\xi\,N_2\,\sqrt{\Om\rho}\,\mathrm{J}_{\frac{1}{1-\eps}}\left(\frac{\rho}{\Om}\right)^{\frac{\eps-1}{2}}
\end{equation}
with $N_2=-2^\frac{1}{1-\eps}\,\Gamma\big(\frac{1}{1-\eps}\big)/(1-\eps)$.

This spacetime has infinitely many Killing horizons and a curvature singularity at $\rho=0$.
The scalar field  is given by a remarkably simple expression,
\begin{equation}
\phi(\rho)=\Big(\frac{\rho}{\Om}\Big)^\frac{\eps-1}{2}\,,
\end{equation}
which allows a straightforward inversion to extract $\rho(\phi)$.
The expression for the potential for different values of $\eps$ is lengthy and not too illuminating, but can be obtained analytically. For example, if $\eps=\tfrac23$ and $\xi=0$ we obtain
\begin{multline}
V(\phi) = 2 + {{\pi^2\phi^4}\over 18432} \big[2 \phi^2 ( \phi^2 - 54)\,\textrm{Y}_1^2(\phi) - 
     4 \phi (13 \phi^2 - 216) \,\textrm{Y}_1(\phi) \textrm{Y}_2(\phi) \\
- (3\phi^4 -176 \phi^2 + 1728)  \,\textrm{Y}_2^2(\phi) + 2 \phi^4 \textrm{Y}_3^2(\phi) - \phi^4 \textrm{Y}_4^2(\phi)\big]
\label{eq:BesselV}
\end{multline}
As expected, the potential does not depend on $\Om$.
Note that there is a subtlety in obtaining the full potential as a function of the scalar field for non-integer $\frac{1}{1-\eps}$, because the potential becomes complex for $\phi<0$. Remember that we fixed a sign ambiguity  in (\ref{eq:quench13}) by assuming that $\dot{\phi}$ is non-positive.
By allowing $\dot{\phi}$ also to be positive we can glue together the two branches and obtain the full (real) potential.

\begin{figure}
\centering
\includegraphics[width=7cm]{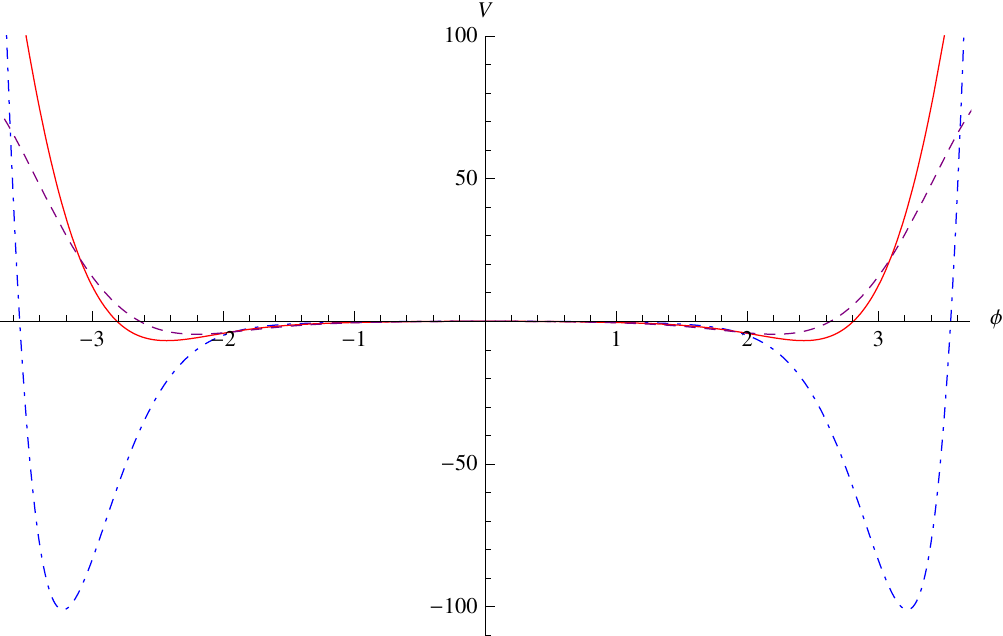}
\includegraphics{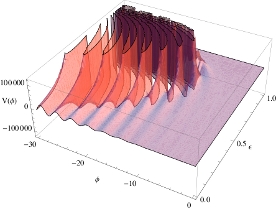} 
\caption{Left plot: potential $V(\phi)$ for $\eps=\tfrac13$ (purple-dashed), $\eps=\tfrac12$ (red-solid) and $\eps=\tfrac23$ (blue dotted-dashed), Right plot: $V(\phi)$ for $\eps\in(0,\,1)$ and large negative values of $\phi$.}
\label{Besselpot}
\end{figure}

In figure \ref{Besselpot} we plot the potential for three different values of $\eps$ and also show the oscillatory behavior in the deep infrared (large values of $\phi$) which originates from the behavior of the Bessel functions.

\subsection{Monomial Ansatz --- solutions with exponential potential}\label{se:2.5}

Here we study a simple case with $V\neq 0$ that is neither asymptotically flat nor asymptotically (A)dS, just to demonstrate that such solutions exist and can be constructed as well following our algorithm. They are of little physical interest, though.

Consider the power-law Ansatz
\eq{
f_1 = 2\rho\, (\rho/\Om)^{\al-1}\qquad 0 < \alpha < 1
}{eq:quench55}
with some $\Om\neq 0$.
Then $f_2$ (as long as $\al\neq\tfrac12$) is determined as
\eq{
f_2 = \chi\,f_1+\frac{\xi\,\Om}{1-2\al}\,(\rho/\Om)^{1-\al}
}{eq:quench56}
where $\chi$ is some constant.
The scalar field $\phi$ is given by
\eq{
\phi = \phi_0 \pm \sqrt{\al(1-\al)}\,\ln\frac{\rho}{\Om}\,.
}{eq:quench57}
The potential is then given by
\eq{
V(\phi;\,\al,\,v_0) = 2 - v_0\,e^{\mp 2\phi\,\sqrt{1/\al-1}}\,.
}{eq:quench58}
Note that $V$ is independent from the constants of motion $j=\Om\,\xi$ and $\Om$. Upon redefining $\phi_0$ also the constant $\chi$ can be absorbed.
The constants $\al$ and $v_0(\al,\,\phi_0)$ are then parameters of the model.
No further constants of motion appear in this construction, which means that further solutions to a model with the potential \eqref{eq:quench58} must have non-monomial solutions for $f_1$.
The case $\al=\tfrac 12$ can be treated similarly, leading to the potential \eqref{eq:quench58} with $\al=\tfrac 12$ and to $f_2=f_1\,\big(\chi+\frac \xi2\,\ln(\rho/\Om)\big)$.

Non-trivial solutions with exponential potential always have a scalar field that diverges logarithmically at $\rho\to\infty$ and $\rho\to 0$.

\section{Generalizations}\label{se:6}

Our algorithm to bootstrap gravity solutions by providing geometric input, in the form of one free function $f_1(\rho)$ subject to the inequality \eqref{eq:quench14}, in principle allows to construct all stationary axi-symmetric solutions of all EDG models in three dimensions \eqref{eq:quench1}. Finding all solutions for fixed scalar potential $V$ turned out to be very hard, so we restricted ourselves to a co-dimension 2 family of solutions. 

Here we discuss some possible generalizations of our methods. We start with generalizations within 3-dimensional EDG.
It is straightforward to drop axi-symmetry and replace it by translation invariance, which leaves the local analysis unchanged and just affects some of the global properties. For instance, the hairy black hole solutions in section \ref{se:3} then lead to hairy black brane solutions.
In section \ref{se:2} we have repeatedly assumed certain smoothness conditions, in particular analyticity. It could be interesting to relax these assumptions and check if it leads to qualitatively new solutions. A precedent for this is the numerical evidence for stationary axi-symmetric solitonic solutions of topologically massive gravity that are not analytic \cite{Ertl:2010dh}.
The solitonic solutions discussed in sections \ref{se:2.3} and \ref{se:6.2} are special and can never have a mass below that of AdS for marginal boundary conditions. However, it is known from ``designer gravity'' \cite{Hertog:2004ns} that solitons can exist whose mass is below that of AdS. These solitons are in fact equivalent to our solitons as solutions to the EOM, but they require non-marginal boundary conditions and must have a positive trace of the associated boundary stress tensor. 

Within EDG there are no locally Schr\"odinger or Lifshitz solutions, because in terms of Cl{\'e}ment's Ansatz \eqref{eq:quench4} these solutions require $\vecX^2 \neq 0 \neq\ddot\vecX$ and $\ddot\vecX^2 = 0$ \cite{Clement:1994sb,Ertl:2010dh}. There is no solution to the Einstein equations with these properties, since \eqref{eq:einstein} implies proportionality of $\vecX$ and $\ddot\vecX$. This provides a motivation to consider theories that are more general than EDG. 

It would be of interest to consider Einstein--Maxwell theory along the lines of EDG, since this theory has the same number of physical degrees of freedom as EDG, so one might expect a comparable amount of complexity. The crucial differences to EDG are the absence of some free function (no scalar potential) and the presence of a gauge symmetry. For applications to holographic superconductors \cite{Gubser:2008px,Hartnoll:2008vx} 
it is of interest to combine these two theories and study Einstein--Maxwell theory with a charged scalar field. This may allow to find analytic toy models for holographic superconductors. Adding Chern--Simons terms --- either gravitational ones or for a Yang--Mills field \cite{Deser:1982vy} 
--- provides another interesting generalization that is capable of producing locally Schr\"odinger, Lifshitz and warped solutions.

In higher dimensions, $D\!>\!3$, similar methods can be used, provided one has sufficiently many, $D-1$, mutually commuting Killing vectors available (see \cite{Anabalon:2012ta} for recent examples of hairy black holes in dimensions $D=4,\, 5$). 
Then one can torically reduce the action, derive an analog of the Cl{\'e}ment action \eqref{eq:quench5} and build up an algorithm analog to the one described in section \ref{se:2}.

\acknowledgments

We thank Sabine Ertl, Michael Gary, Niklas Johansson, Karl Landsteiner, Cristian Martinez, German Sierra, Paul Tod and Ricardo Troncoso for discussions.

JA is supported by the Portuguese Funda\c{c}{\~a}o para a Ci{\^e}ncia e Tecnologia, grant SFRH/BD/45988/2008.
DG and SS are supported by the START project Y435-N16 of the Austrian Science Fund (FWF) and by the FWF project P21927-N16.
EL has been supported by grants FPA-2009-07908, HEPHACOS S2009/ESP-1473 and CPAN (CSD2007-00042).
IP is funded by the JAE-DOC program of the Spanish research council under the contract JAEDOC068.
The authors acknowledge the support of the Spanish MINECOs `Centro de Excelencia Severo Ochoa' Programme under grant SEV--2012--0249.


\providecommand{\href}[2]{#2}\begingroup\raggedright\endgroup

\end{document}